\documentclass[preprint,12pt]{elsarticle}

\usepackage[utf8]{inputenc}
\usepackage{graphicx}
\usepackage{subfigure}
\usepackage{caption}
\usepackage{threeparttable}
\usepackage{color}
\usepackage{multirow}
\usepackage{lineno,hyperref}
\usepackage[bottom]{footmisc}
\modulolinenumbers[5]
\usepackage{rotating}
\usepackage{changes}
\usepackage{times}
\journal{Online/Mobile Social Networking at the time of COVID-19}

\begin{document}

\begin{frontmatter}

\title{An Exploratory Study of COVID-19 Information on Twitter in the Greater Region}

\author[ad1]{Ninghan Chen}
\ead{ninghan.chen@uni.lu}
\author[ad1]{Zhiqiang Zhong}
\ead{zhiqang.zhong@uni.lu}
\author[ad1,ad2]{Jun Pang\corref{cor1}}
\ead{jun.pang@uni.lu}
\address[ad1]{Faculty of Sciences, Technology and Medicine,
University of Luxembourg, L-4364 Esch-sur-Alzette,
Luxembourg}
\address[ad2]{Interdisciplinary Centre for Security, Reliability and Trust,
University of Luxembourg, L-4364 Esch-sur-Alzette,
Luxembourg}
\cortext[cor1]{Corresponding author}

\addtolength{\abovedisplayskip}{-1ex}
\addtolength{\belowdisplayskip}{-1ex}

\pagestyle{plain}

\begin{abstract}
The outbreak of the COVID-19 leads to a burst of information in major online social networks (OSNs). Facing this constantly changing situation, OSNs have become an essential platform for people expressing opinions and seeking up-to-the-minute information. Thus, discussions on OSNs may become a reflection of reality.
This paper aims to figure out the distinctive characteristics of the Greater Region (GR) through conducting a data-driven exploratory study of Twitter COVID-19 information in the GR and related countries using machine learning and representation learning methods. We find that tweets volume and COVID-19 cases in GR and related countries are correlated, but this correlation only exists in a particular period of the pandemic. Moreover, we plot the changing of topics in each country and region from 2020-01-22 to 2020-06-05, figuring out the main differences between GR and related countries.

\end{abstract}


\end{frontmatter}


\section{Introduction}
\label{sec:intro}

The outbreak of the COVID-19 leads to an outbreak of information in major online social networks (OSNs), including Twitter, Facebook, Instagram, and YouTube~\cite{cinelli2020covid}.
Facing this massive COVID-19 outbreak and constantly changing situation, OSNs have become an essential platform for people to seek up-to-the-minute and local information.  Moreover, due to physical isolation and social distancing, people spend much more time on OSNs — engaging in expressing opinions, encouraging others, openly lambasting mismanagement, and voicing vitriol, etc. Discussions on OSNs can be a reflection of reality, and topics about the pandemic mirror the public concerns in real-time.
A growing number of research links OSNs activities to COVID-19. Existing literature has already demonstrated that posts about the pendemic on OSNs can be a leading indicator of COVID-19 daily cases~\cite{singh2020first,jahanbin2020using}, related discussions on OSNs can be categorised into multiple specific topics~\cite{wang2011collaborative,ordun2020exploratory,medford2020infodemic,sharma2020covid}, and OSNs may help to design more efficient pandemic models for social behaviour and thus the government can implement more responsive communication strategies~\cite{cinelli2020covid,gupta2020tracking,bento2020evidence}. 
However, there are three main problems within the existing researches.
First, researches with geographic data are based on coarse-grained processing of the location information~\cite{singh2020first,lopez2020understanding}.
Second, the existing topic modelling studies mostly focus on different topics in a relatively long period (weeks or months)~\cite{medford2020infodemic,cinelli2020covid} and general characteristics of user concerns, which cannot provide a precise representation of how topics change on a daily basis.
Third, shared information on OSNs over the global or nations~\cite{thelwall2020retweeting,singh2020first,lopez2020understanding} are too general in terms of geographic dividing.

When analysing the COVID-19 information on Twitter by geographic locations, it cannot be ignored that the movement of population shapes the spatio-temporal patterns of the pandemic~\cite{balcan2009multiscale}. Population mobility plays an important role in the spread of COVID-19. In other words, in terms of regions with highly frequent and mobile cross-border commuters, researches only concerning political sovereign states are biased.

To fill this gap, we introduce the concept of the `relational city'. Relational city is defined as a region that ``constituted through globally critical flows of capital, goods, and ideas, and whose economies are dedicated to intermediary services such as offshore banking, container- and bulk-shipping, and regional re-exportation''~\cite{sigler2013relational}. in a nutshell, relational city is a specific set of cities that exhibit spatial transformations due to the influence of advanced capitalism, and it can be transnational.
To be more specific, these cities tend to be located in cross-border regions, influenced by different linguistic, cultural, and political systems, and rely heavily on exchange economy, with a large number of cross-border workers. The high mobility of cross-border workers brings high risks of virus spreading. Studies have shown that lockdown in relational cities is likely to have more severe impact on economy than in cities of traditional concept~\cite{hesse2020relational}. However, up to the present time, there is no data-driven analysis of OSNs COVID-19 information about any relational city yet. The Greater Region (GR), a typical relational city with Luxembourg at its centre and adjacent regions of Belgium, Germany and France (i.e., Wallonia, Saarland, Lorraine, Rhineland-Palatinate and the German-speaking Community of Belgium) is chosen as the representative in our case study. We define the countries mentioned above as the related countries of GR.

GR has the highest number of cross-border commuters in Europe, approximately $250,000$ per day.\footnote{https://bit.ly/2P6NLSm}
This makes GR a particular and outstanding example: virus  spreads due to its high mobility, as the whole business model in GR requires a large number of cross-border workers to sustain.  With the implementation of a set of policies including border closures and the progression of the pandemic, GR is affected in economy, daily life, travel, and other aspects.

This study focuses on two dimensions, tweet volume (see Section~\ref{sec:q1})  and tweet text (see Section~\ref{sec:topics}) to analyse Twitter information in GR and related countries about COVID-19. The following two main questions are addressed in the corresponding section.

\begin{enumerate}
\item[{\sf RQ1}] Whether there is a strong correlation between tweet volume and COVID-19 daily cases in GR and related countries, and, if so, whether tweet volume can help predict COVID-19 daily cases?
\item[{\sf RQ2}] Whether there are distinctive characteristics of these region and countries’ topics about COVID-19 on Twitter, and whether GR, as a relational city, embodies any characteristics in the topics?
\end{enumerate}

We collected $51,966,639$ tweets from Twitter, which are posted by $15,551,266$ Twitter users all over the world from 2020/01/22 to 2020/06/05. Among them are $1,643,308$ posts posted by $41,690$ users in GR and its related countries. To investigate {\sf RQ1}, 
basic reproductive rate $R_0$ and effective reproductive rate $R(t)$ in epidemiology~\cite{heesterbeek1996concept} are introduced to slice the pandemic periods, and correlations between tweet volume and daily cases in each period are calculated by Pearson Correlations (PC). 
A novel topic modelling method combing Bidirectional Encoder Representations from Transformers (BERT)~\cite{devlin2018bert} and the Latent Dirichlet Allocation (LDA) topic modelling method~\cite{blei2003lda} is introduced, and a supervised Support Vector Machine (SVM)~\cite{chang2011libsvm} for classifying topics into given categories is trained to study {\sf RQ2}.

The main contributions in this paper are threefold.
\begin{enumerate}
\item[(I)] We screen a novel Twitter dataset of 2020/01/022 to 2020-06-05 which contains data from users with locations labelled in GR, and related countries including Luxembourg, France, Germany and Belgium, and the COVID-19 related tweets from Chen et al’s dataset~\cite{COVID-19Dataset}. This dataset will be shared with the public to advance related research.

\item[(II)] Spatio-temporal analysis is carried out to showcase how the COVID-19 daily cases are correlated with tweet volume in a long period. We find that tweet volume and COVID-19 daily cases in GR and related countries are correlated, and tweet volume can help predict COVID-19 daily cases, but this strong correlation only exists during the early period of the pandemic.

 \item[(III)]We find that GR, as a relational city, has distinctive characteristics in the topics. Users in GR show more concerns in anti-contagion and treatment measures before COVID-19 reaches its peak, and have a higher level of interest in policy and daily life before $R(t) < 1$ than the related countries.

\end{enumerate}

This study sheds light on how the Twitter users in GR and related countries react differently over time through an interdisciplinary approach. It may, therefore, help to understand changes in public concerns on Twitter during the pandemic, and in particular, the distinctive characteristics of topics in GR, a relational city with high mobility.

\section{Related Work}
\label{sec:Related work}

Some existing results have already shown that social media conversations can be a leading predictor of a new pandemic cases~\cite{singh2020first, st2012can,jahanbin2020using}, and in many countries tweets increase in volume before the number of conﬁrmed cases increases. Studies have shown that anti-contagion policies can significantly and substantially reduce the spread of COVID-19~\cite{hsiang2020effect, courtemanche2020strong,dergiades2020effectiveness}, and the effect of policies on the mitigation of spread varies, influenced by factors including culture, demographic information, socio-economic status and national health systems, where changes in public knowledge may affect the impact of the policies. If the public adjusts their behaviour in response to information from sources that are not policy-related, it may change the spread of COVID-19~\cite{hsiang2020effect}. 

Researches of public behaviour patterns of the pandemic have been conducted based on data from smart devices~\cite{gupta2020tracking}, search index~\cite{hu2020more,effenberger2020association}, and COVID-19 related conversations on Twitter. Bento et al.~\cite{bento2020evidence} mention that, there is a spike in searches for basic information about Covid-19 when the first case was announced in each state in the United States, but the first case report does not trigger discussions about policy and daily life. Topic modelling, an unsupervised approach that detects latent semantic structure~\cite{wang2011collaborative} is widely used. Cinelli et al.~\cite{cinelli2020covid} extract topics with word embedding on a global scale, making the conclusion that social media may help to design more efficient epidemic models for social behaviour and to implement more timesaving communication strategies. The LDA model is used by Medford et al.~\cite{ordun2020exploratory} and Ordun et al.~\cite{medford2020infodemic} to analyse the topics in the early period of the pandemic. Sharma, et al.~\cite{sharma2020covid} use character embedding~\cite{joulin2016fasttext} 
and Term Frequency Inverse Document Frequency (TF-IDF) word distribution with manual inspection for topic modelling. However, LDA, a bag-of-words approach, which is widely used to identify latent subject information in a large-scale document collection or corpus, has some drawbacks: it needs large corpus to train, ignores contextual information and performs mediocrely in handling short texts~\cite{yan2013biterm}. As a result, these studies extract the topic over certain time periods, and the time granules are too coarse to accurately reflect the trend of the topics. 

\section{Data Description}
\label{sec:dataset}
In this section, we briefly describe how we screened COVID-19 tweets from Chen et al.'s dataset~\cite{COVID-19Dataset} to build our dataset of GR and the related countries, and how we obtained information on COVID-19 daily cases for these region and countries.

\subsection{Twitter data collection}
Twitter, one of the most prominent online social media platform, has been used extensively during the pandemic. In this study, $51,966,639$ tweets posted by more than $15$ million Twitter users from 2020/01/22 to 2020/06/05 are hydrated from Chen et al.'s dataset~\cite{COVID-19Dataset} via the Twitter Streaming API.
This COVID-19 twitter dataset collects tweets with specific keywords including `COVID-19', `coronavirus', `lockdown', etc. Attribute with * in Table~\ref{table:sample_dataset} is contained in the dataset.
To comply with Twitter’s Terms of Service, they only publicly released the tweet ids of the collected tweets.
To compose our dataset, we first hydrated raw data via the API based on the tweet ids they provided, which included tweet id, full text, user id and user-defined location information.

\begin{figure}[]
\centering 
\subfigure[GR]{
\label{Fig.sub.1}
\includegraphics[width=0.45\textwidth]{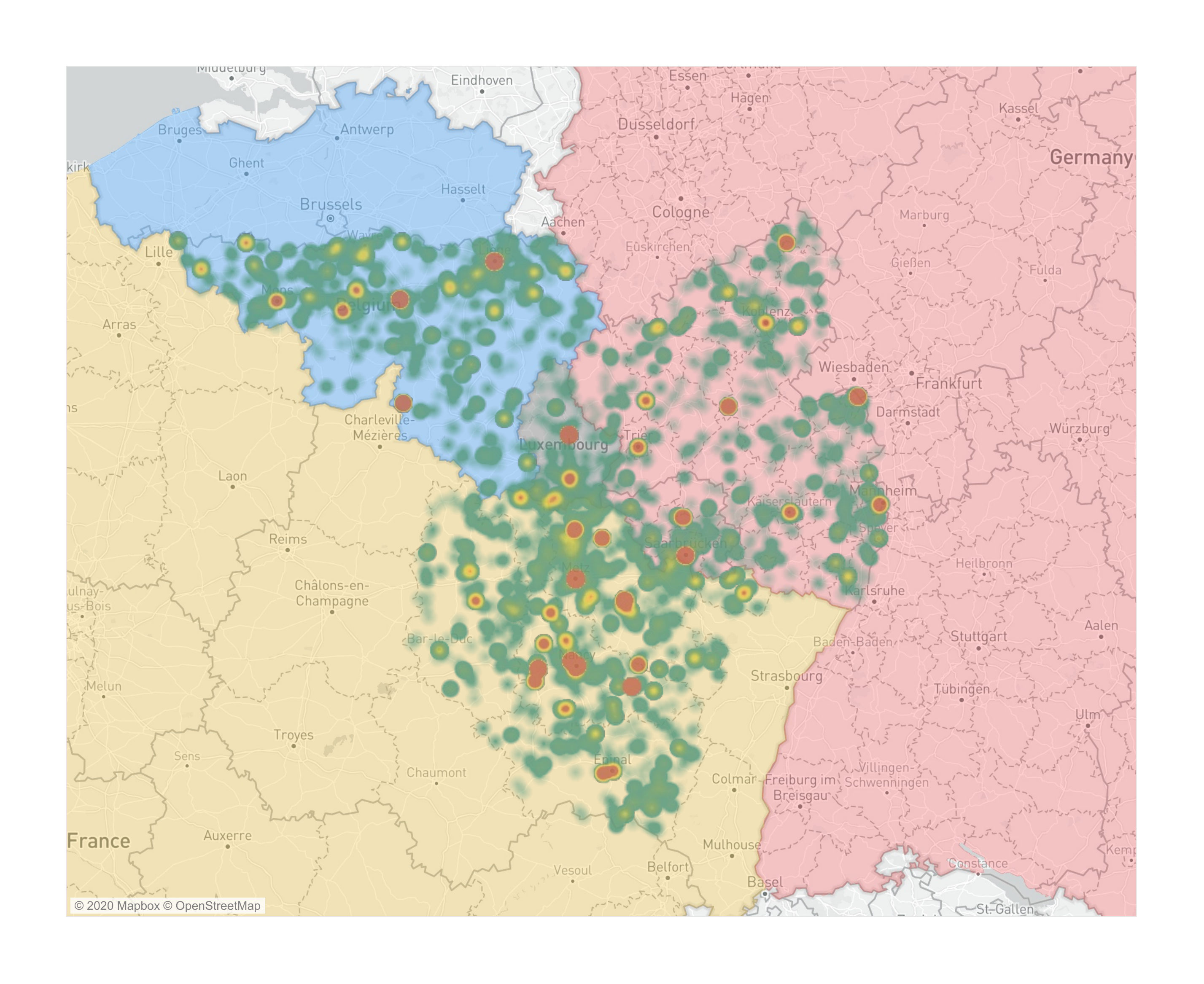}}
\subfigure[The related countries]{
\label{Fig.sub.2}
\includegraphics[width=0.45\textwidth]{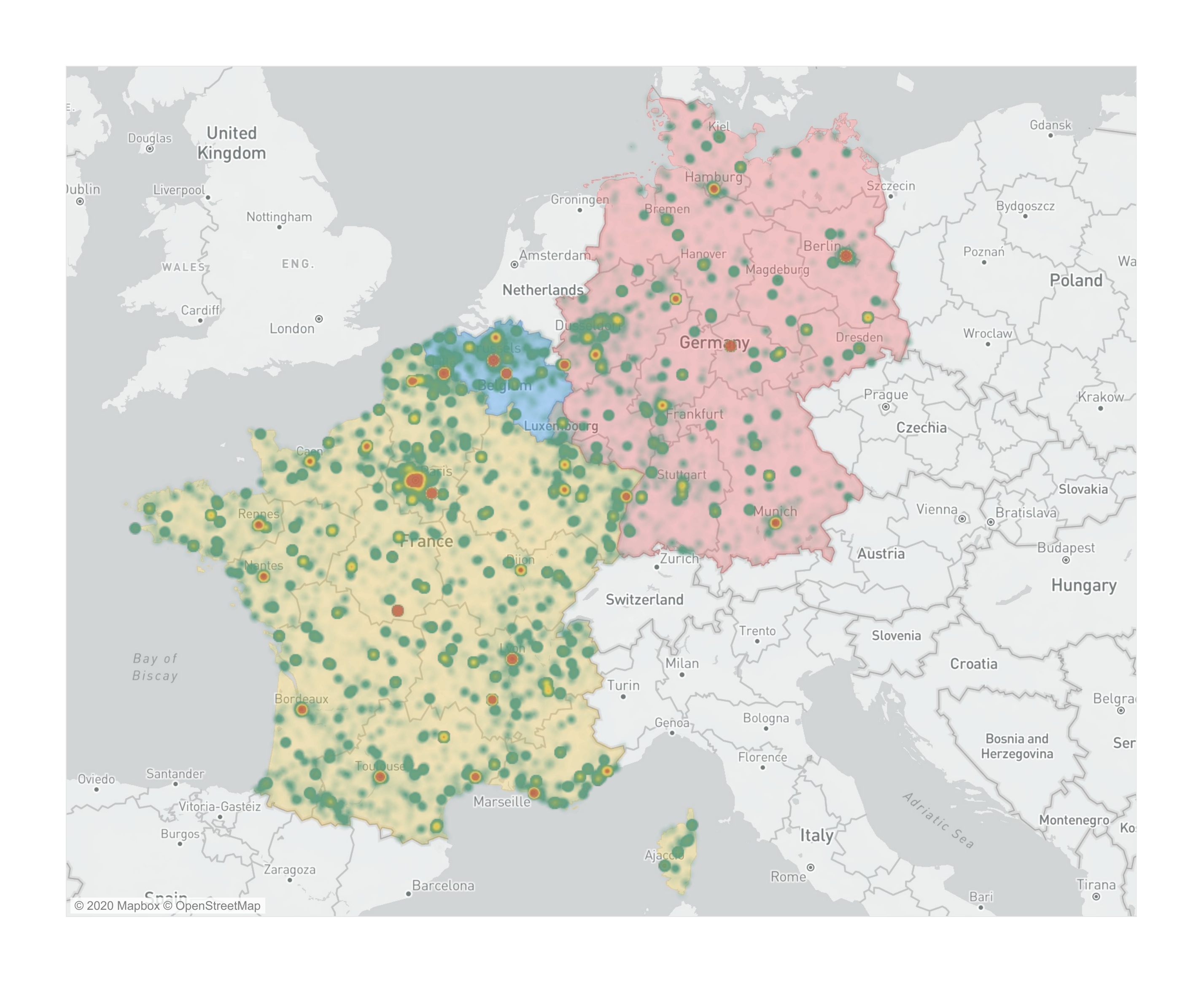}}
\caption{User location heatmap of GR and the related countries}
\label{fig:map}
\end{figure}

Secondly, as the user location information we collected so far is user-defined, nether accurately revealing a true location nor machine-parseable, we processed the fuzzy location context into real location information by leveraging geocoding APIs, Geopy\footnote{\url{https://bit.ly/3gfW2PP}} and
ArcGis Geocoding\footnote{\url{https://bit.ly/3f9OUDa}}. 
In more detail, user-defined locations in many cases, detailed country locations are not included, usually just a city or an abbreviation of a state. If user-defined locations are matched directly based on characters, users who fill in this kind of context will be ignored.
Geopy, a Python client for geocoding services and ArcGIS Geocoding, a geographic information services system, geocode a fuzzy string into a complete address of a fixed format including state and country. For example, `Moselle' in Table~\ref{table:sample_dataset} would be geocoded as `Moselle, Lorraine, France'. Once the location information has been geocoded, users located in the GR, Luxembourg, France, Germany, and Belgium are screened by character matching. Table~\ref{table:sample_dataset} gives an example in the final dataset, and Table~\ref{table:summary_dataset} shows the summary of the collected tweet data of GR, Luxembourg, France, Germany, Belgium and the global. Figure~\ref{fig:map} contains two heatmaps of user location in GR and the related countries for a better understanding of this study.

\begin{table}[!h]
\centering
\resizebox{\textwidth}{!}{%
\begin{tabular}{|l|l|l|}
\hline
Attribute & Description & Example \\ \hline
Tweet\_id\added{*} & A unique identifier for a Tweet & 12319668395****** \\ \hline
Full\_text& Text of a tweet & \begin{tabular}[c]{@{}l@{}}RT @******:\\
The Diamond princess is\\
a UK ship managed by the US. \\ UK should Be Responsible.\\
\#DiamondPrincess
\#coronavirus\end{tabular} \\ \hline
User\_id& Unique identifier for this user & u9181074902***** \\ \hline
User\_geo\_orginal & User-defined location information & Moselle \\ \hline
User\_geo & Geocoded user location & Moselle, Lorraine, France \\ \hline
\end{tabular}%
}
\caption{A sample of our COVID-19 Twitter dataset}
\label{table:sample_dataset}
\end{table}

\begin{table}[!h]
\centering
\begin{tabular}{|l|r|r|}
\hline
Region/Country & \multicolumn{1}{l|}{tweet volume} & \multicolumn{1}{l|}{User volume} \\ \hline
Global     & 51,966,639 & 15,551,266 \\ \hline
GR         & 35,329     & 7,894      \\ \hline
Luxembourg & 7,512      & 1,545      \\ \hline
Belgium    & 119,467    & 31,446     \\ \hline
France     & 1,050,312  & 288,009    \\ \hline
Germany    & 430,688    & 87,796     \\ \hline
\end{tabular}
\caption{Summary of our COVID-19 Twitter dataset}
\label{table:summary_dataset}
\end{table}

\begin{figure}[!t]
\centering
\begin{minipage}{0.90\textwidth}
\includegraphics[width=0.95\textwidth]{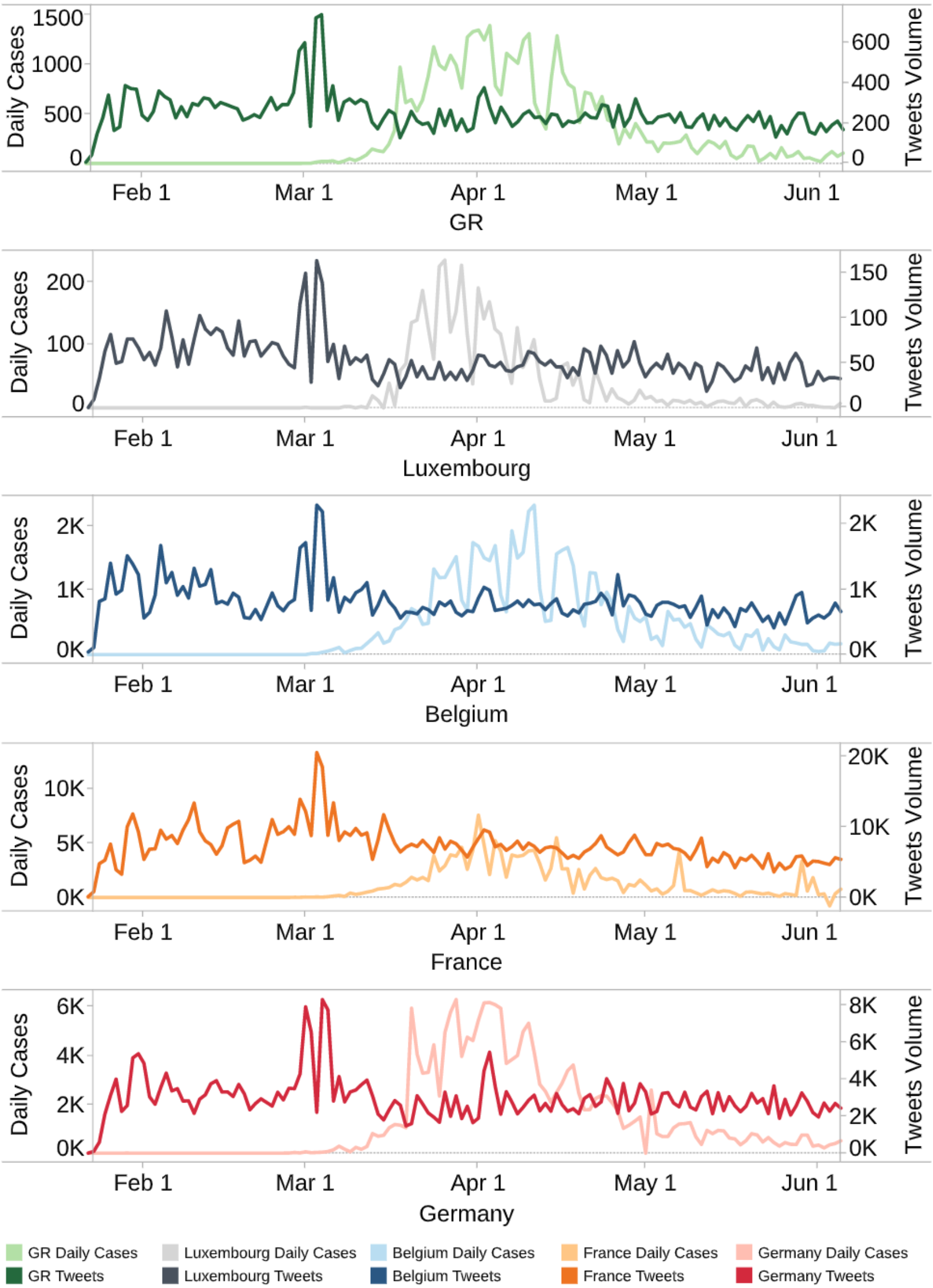}
\caption{Daily tweet volume and COVID-19 new cases\protect\footnote{On 3rd June, France published a revision of data that lead to a negative number of new cases, see \url{https://bit.ly/33c8CM8} for the original news.}}
\label{fig:daily tweets}
 \end{minipage}
\end{figure}
\subsection{COVID-19 data collection}

The dataset published by the European Center for Disease Prevention and Control\footnote{\url{https://bit.ly/3jYhefx}} allows us to obtain COVID-19 data including daily cases, deaths and locations for the country we selected.
As there is no official COVID-19 data published for GR, which is composed of Luxembourg, Wallonia in Belgium, Saarland and Rhineland-Palatinate in Germany and Lorraine in France, we add up all the data for the cities and regions mentioned above from the datasets\footnote{\url{https://bit.ly/2ErDii7,https://bit.ly/3gaGGMm,https://bit.ly/33c8CM8}} 
published by corresponding countries as the final GR data when counting daily cases and deaths in the GR. It should be noted that as the number of daily new cases in France is not available at the regional level, and deaths, hospitalisations, departures data have been published only since March 18, 2020, data for Lorraine is counted as zero until March 18, 2020, and the sum of hospitalisations, 
hospital departures and deaths is considered as the total number of cases on that particular day.

\section{Correlation between COVID-19 daily cases and tweet volume}                       
\label{sec:q1}

To explore the correlation between tweet volume and COVID-19 daily cases in GR and the related countries, we introduce basic reproductive rate $R_0$ and effective reproductive rate $R(t)$ in epidemiology to slice the periods of the pandemic, and a spatio-temporal analysis of the correlation between tweet volume and daily cases in each period is conducted by Pearson Correlations ($PC$).

\subsection{$R(t)$-based time division}
\begin{figure}[!t]
    \centering
     \includegraphics[width =\textwidth]{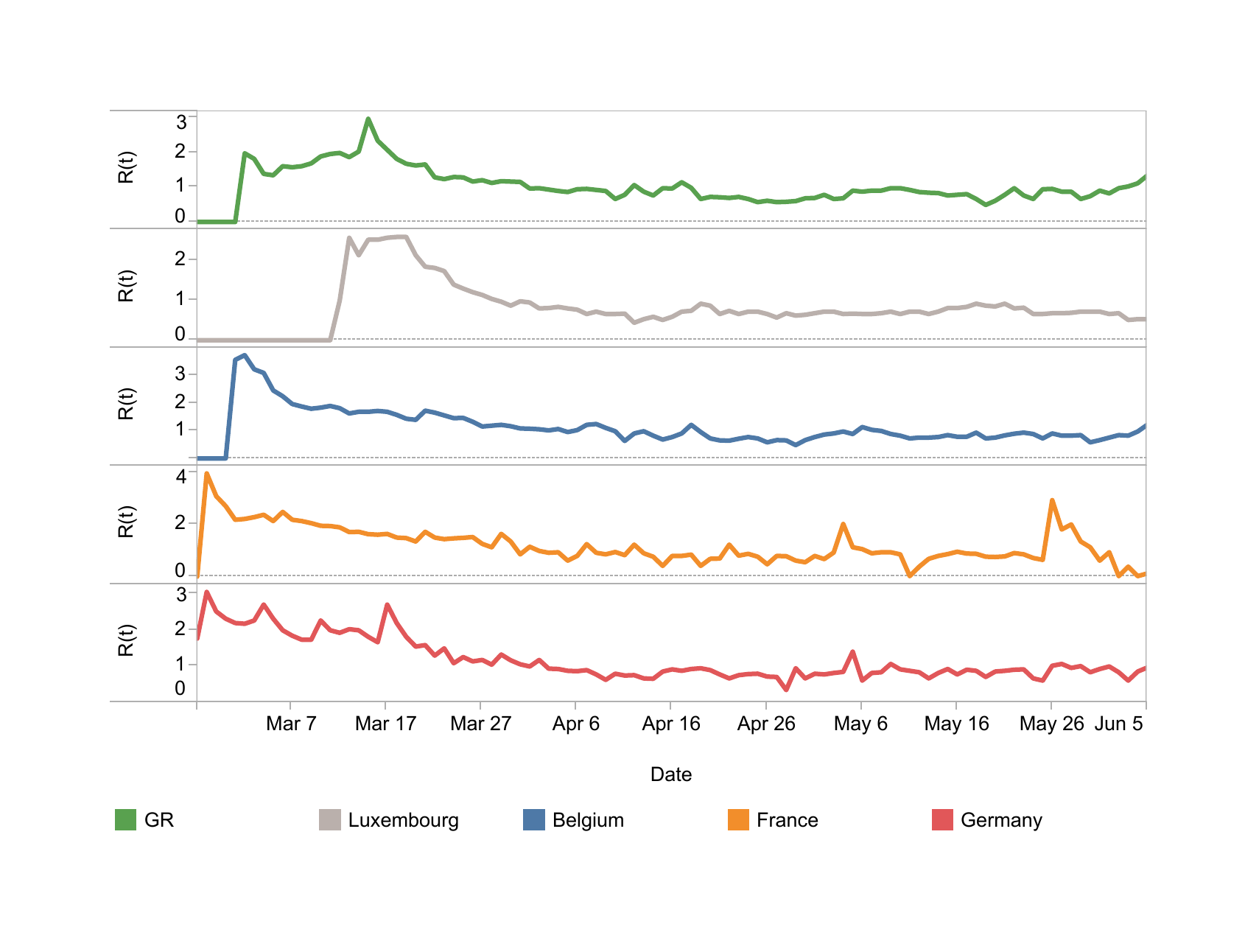}
\vspace{-15mm}
    \caption{Effective reproductive rate ($R(t)$)}
    \label{fig:Rt}
\end{figure}

$R_0$ is the expected number of cases arising directly from a single case in a population where all individuals are susceptible to infection~\cite{heesterbeek1996concept} and $R(t)$ represents the average number of new infections caused by an infected person at time $t$. If $R(t) > 1$, the number of cases will increase, e.g. at the beginning of an epidemic. When  $R(t) = 1 $, the disease is endemic, and when  $R(t) < 1 $, the number of cases will decrease. 
For the calculation of real-time $R(t)$, we use a Bayesian approach~\cite{bettencourt2008real} with Gaussian noise to calculate the time-varying $R(t)$ based on daily new cases, which is also the official method for calculating $R(t)$ in Luxembourg.\footnote{https://github.com/k-sys/covid-19/}
In this case, while the study of calculating $R_0$ of COVID-19 is still ongoing, we use the $R_0$ estimated by WHO\footnote{\url{https://bit.ly/3fgOQkY}}
which $1.4\leq R_0 \leq 2.5$.
The results of time-varying $R(t)$ for GR, Luxembourg, Belgium, France, and Germany are shown in Figure~\ref{fig:Rt}.

The relationship between $R_0$ and the $R(t)$ indicates the spreading ability of the virus. When $R(t)>max(R_0)$, it indicates that the virus is spreading at a higher rate than natural transmission, and the number of cases is about to reach a peak. When $ min(R_0)\leq R_t\leq max(R_0)$, the virus spreads with the basic reproductive rate $R_0$, which implies that the effectiveness of the containment measures is not yet reflected in $R(t)$.
In short, the virus is still spreading freely at its natural transmission.
When $1\leq R_t<min(R_0)$, it means that the virus is spreading at a rate lower than $R_0$, the transmission is impeded, and the containment measures are in effect. When $R_t< 1$, the virus spreads slowly, and can eventually die out.

Here, we divide the pandemic into four periods based on the above analysis, which are: {\it Pre-peak} period (if $ R(t)$ peaks for the first time on day $t_0$ and begins to decrease, with $R(t)< 2.5$ on day $t_1,(t_1\geq t_0)$, then the pre-peak period is the 30-day period before $t_1$).
 {\it Free-contagious} period ($1.4\leq R_t\leq 2.5$); {\it Measures} period ($1\leq R(t)<1.4$); {\it Decay} period ($R(t)<1$). 
It should be noted that the second wave of the pandemic did not begin at the time when this study was conducted, so this division of intervals only applies to this time period, i.e., from 2020-01-22 to 2020-06-05.
The precise time duration of these pandemic periods for each country and region is summarised in Table~\ref{tab:period}.

\begin{table}[!h]
\resizebox{\textwidth}{!}{%
\begin{tabular}{|l|r|r|r|r|}
\hline
 & \multicolumn{1}{c|}{{\it Pre-peak}} & \multicolumn{1}{c|}{{\it Free-contagious}} & \multicolumn{1}{c|}{{\it Measures} period} & \multicolumn{1}{c|}{{\it Decay} period} \\ \hline
GR & 2/14 - 3/15/2020 & 3/15 - 3/21/2020 & 3/21 - 4/17/2020 & 4/17 - 6/05/2020 \\ \hline
Luxembourg & 2/19 - 3/20/2020 & 3/20 - 3/24/2020 & 3/24 - 4/01/2020 & 4/01 - 6/05/2020 \\ \hline
Belgium & 2/04 - 3/05/2020 & 3/05 - 3/25/2020 & 3/25 - 4/18/2020 & 4/18 - 6/05/2020 \\ \hline
France & 2/05 - 3/06/2020 & 3/06 - 3/30/2020 & 3/30 - 4/23/2020 & 4/23 -  6/05/2020 \\ \hline
Germany & 1/29 - 2/28/2020 & 2/28 - 3/24/2020 & 3/24 - 4/02/2020 & 4/02 - 6/05/2020 \\ \hline
\end{tabular}%
}
\caption{Time duration of the four pandemic periods for GR, Luxembourg, Belgium, France and Germany}
\label{tab:period}
\end{table}

\begin{figure}[!t]
    \centering
    \includegraphics[width =\textwidth]{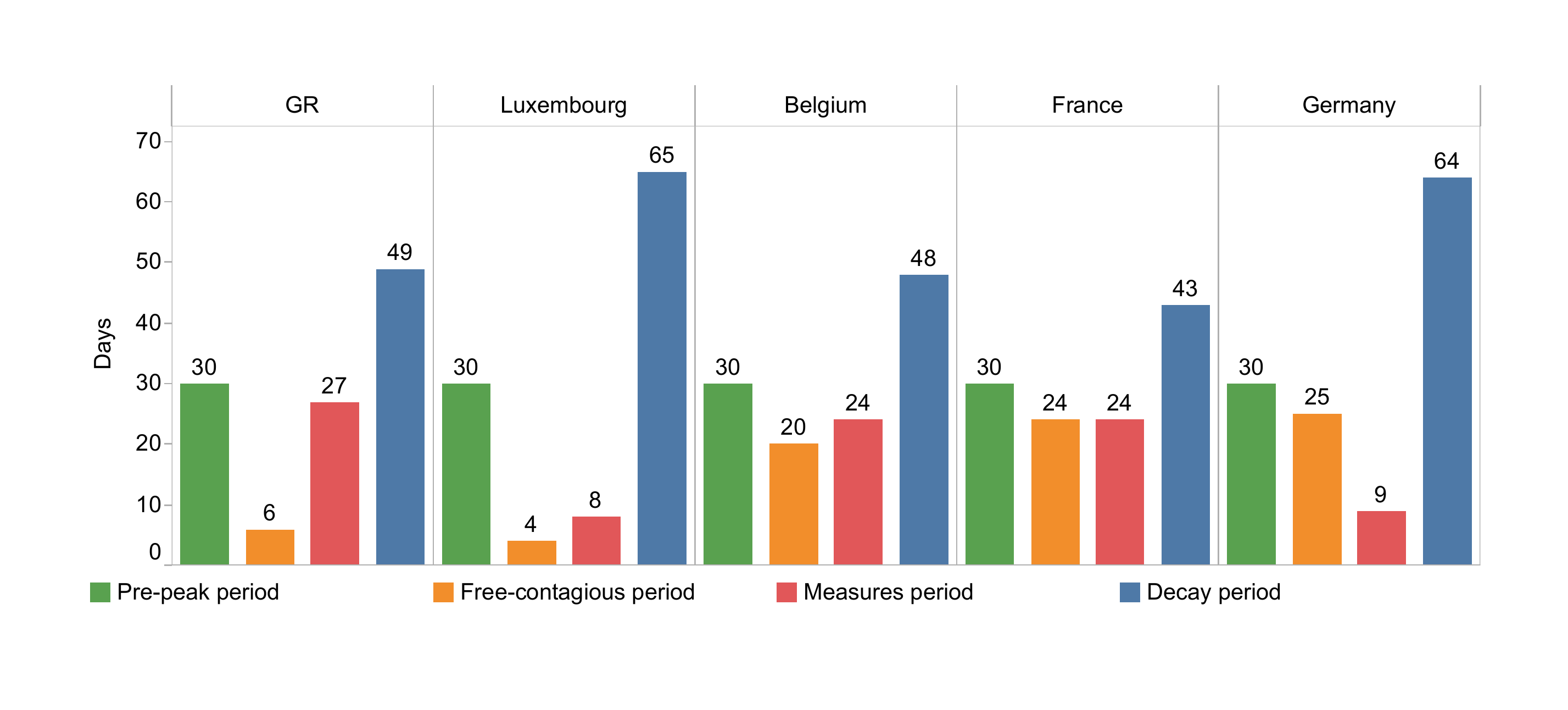}
    \vspace{-15mm}
    \caption{Total days for each pandemic period}
    \label{gra:days}
\end{figure}

The exact numbers of days of each pandemic period are shown in Figure~\ref{gra:days} for the region and countries. The {\it Free-contagious} period in Luxembourg and GR is particularly shorter
(4 \& 6 days)  compared to other countries (24-20 days). 
Being a relational city characterised by high mobility, it may be relatively difficult to control the pandemic. The reason why GR and Luxembourg, has a shorter {\it Free-contagious} period instead, will be discussed in Section~\ref{sec:topics} in terms of the public concerns that reflected by tweet text.


\subsection{Research question {\sf RQ1}}
To answer {\sf RQ1}, we test the following hypotheses:
\begin{enumerate}
\item[{\sf H1}]There is a strong correlation between tweet volume and COVID-19 daily cases in GR and related countries.
\item[{\sf H2}]Tweet volume can help predict COVID-19 daily cases.
\end{enumerate}

\begin{figure*}[!h]
\centering
\subfigure[Pre-peak]{
\begin{minipage}[b]{0.23\linewidth}
\includegraphics[width=1\linewidth]{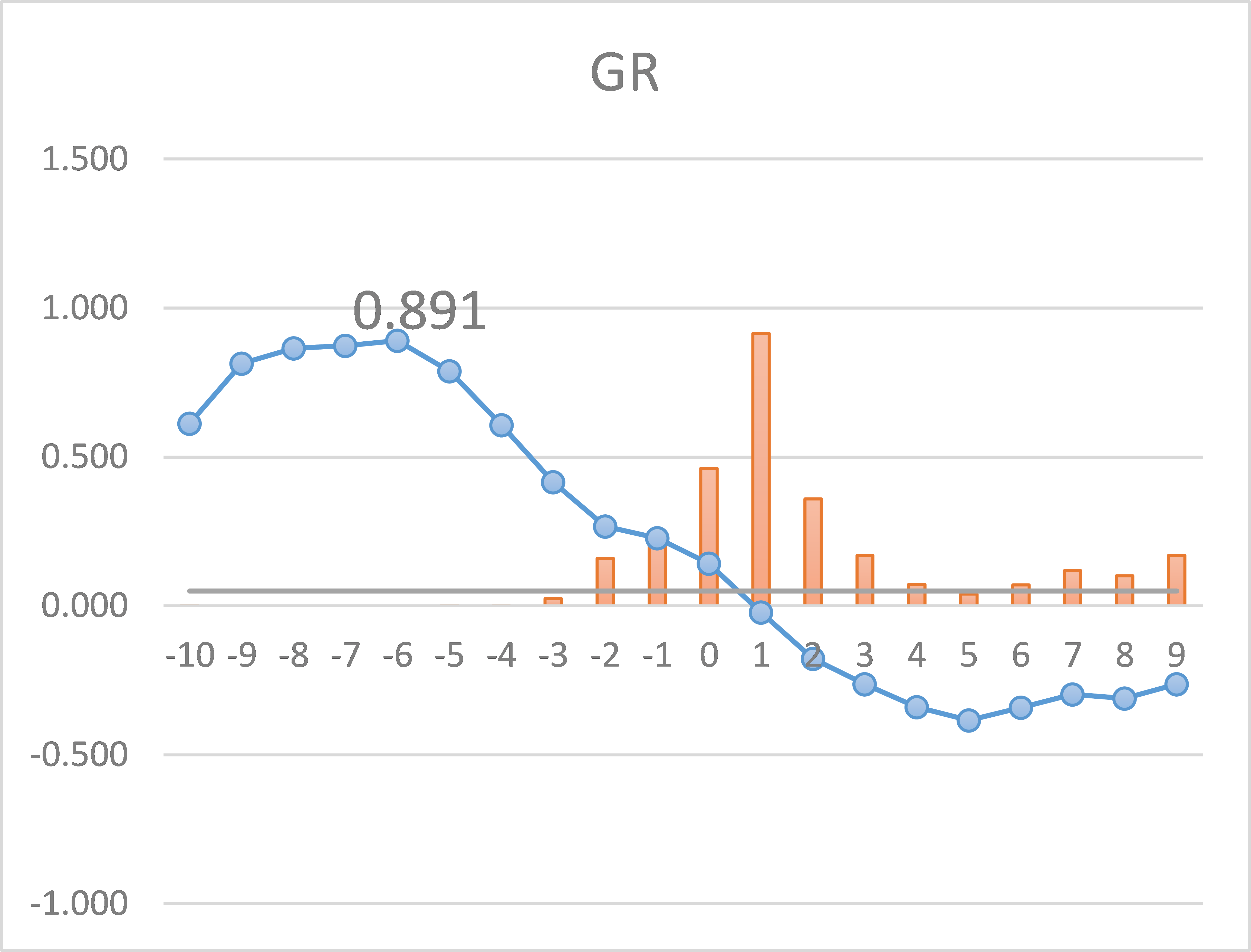}\vspace{0pt}\hspace{0pt}
\includegraphics[width=1\linewidth]{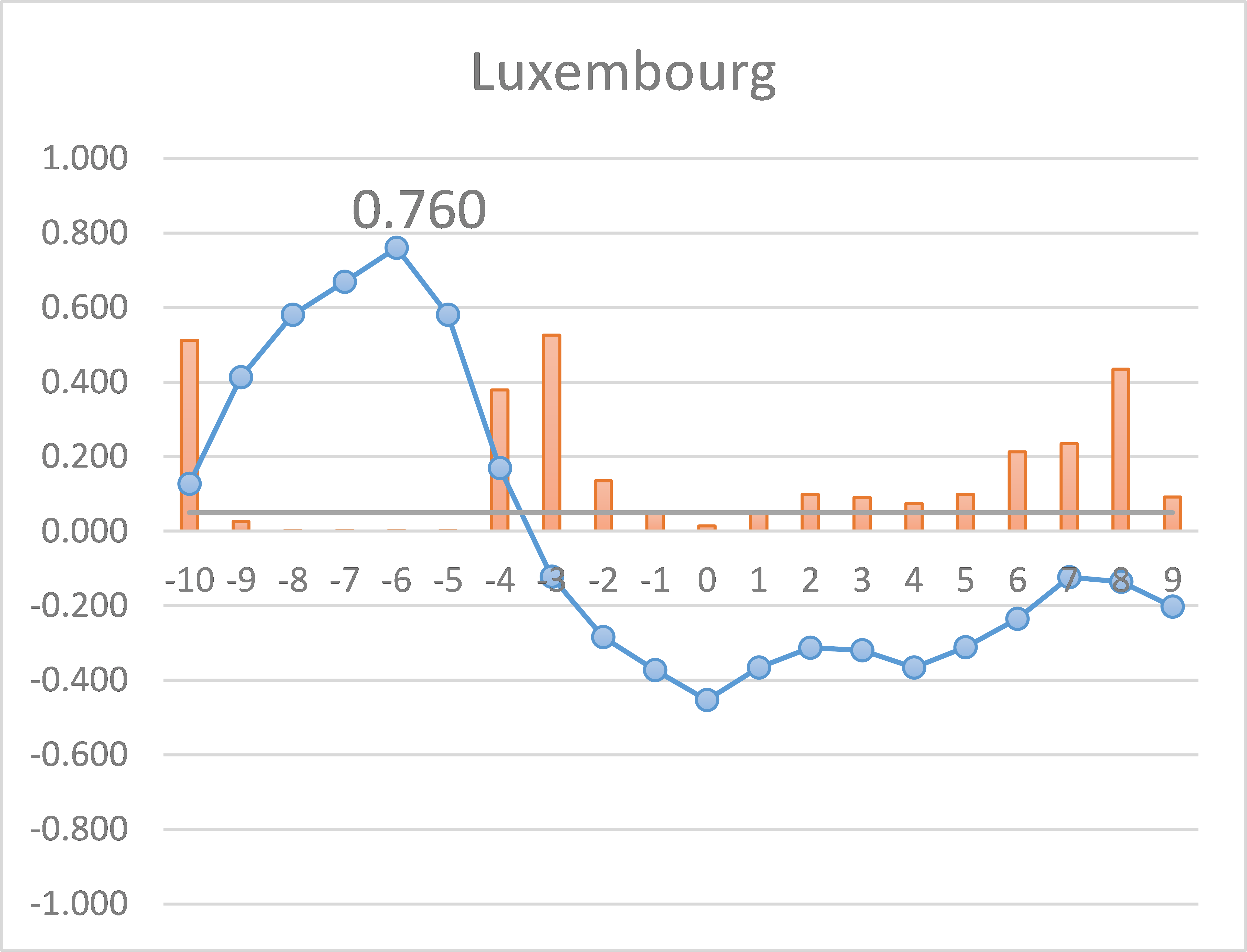}\vspace{0pt}\hspace{0pt}
\includegraphics[width=1\linewidth]{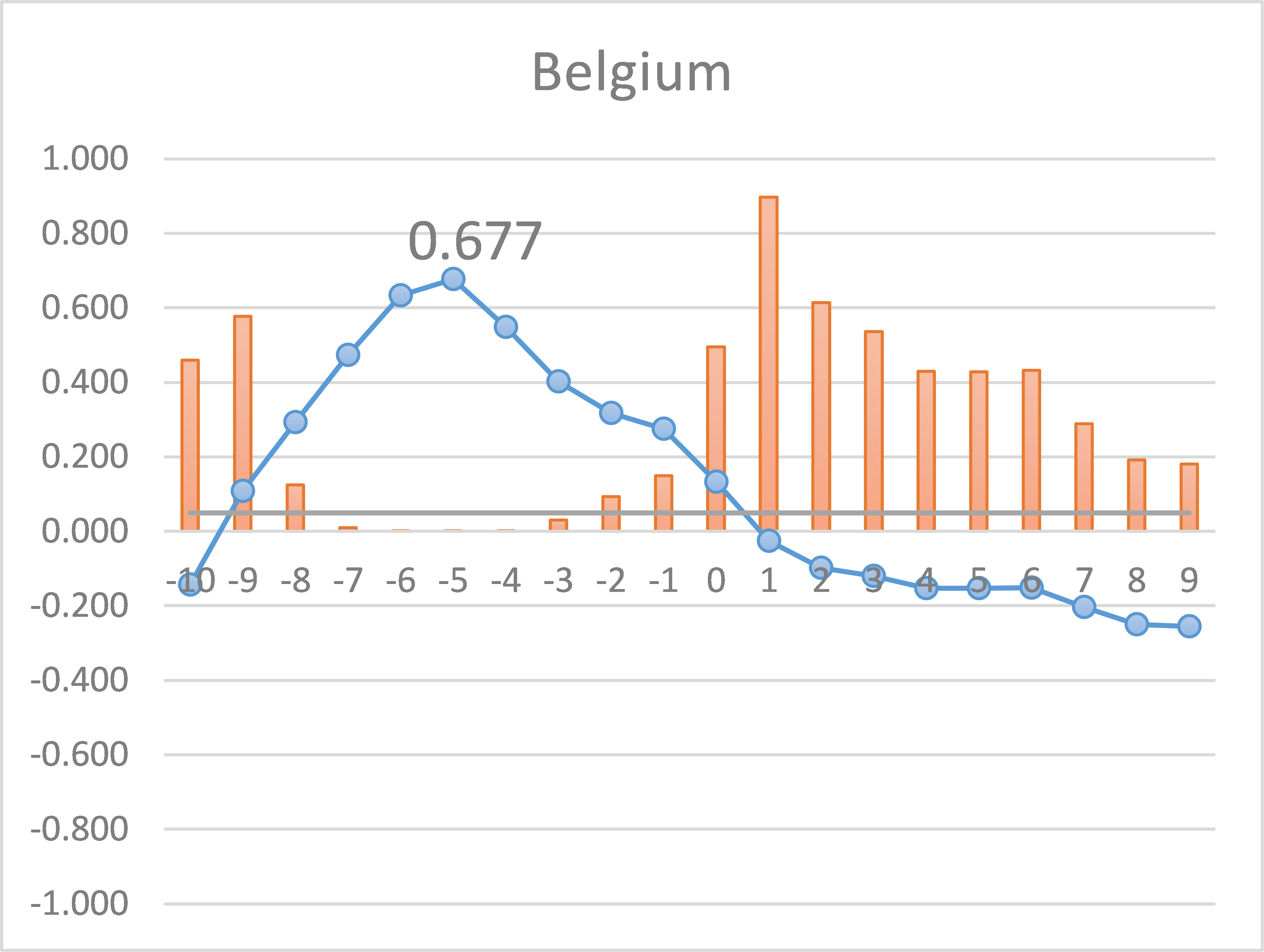}\vspace{0pt}\hspace{0pt}
\includegraphics[width=1\linewidth]{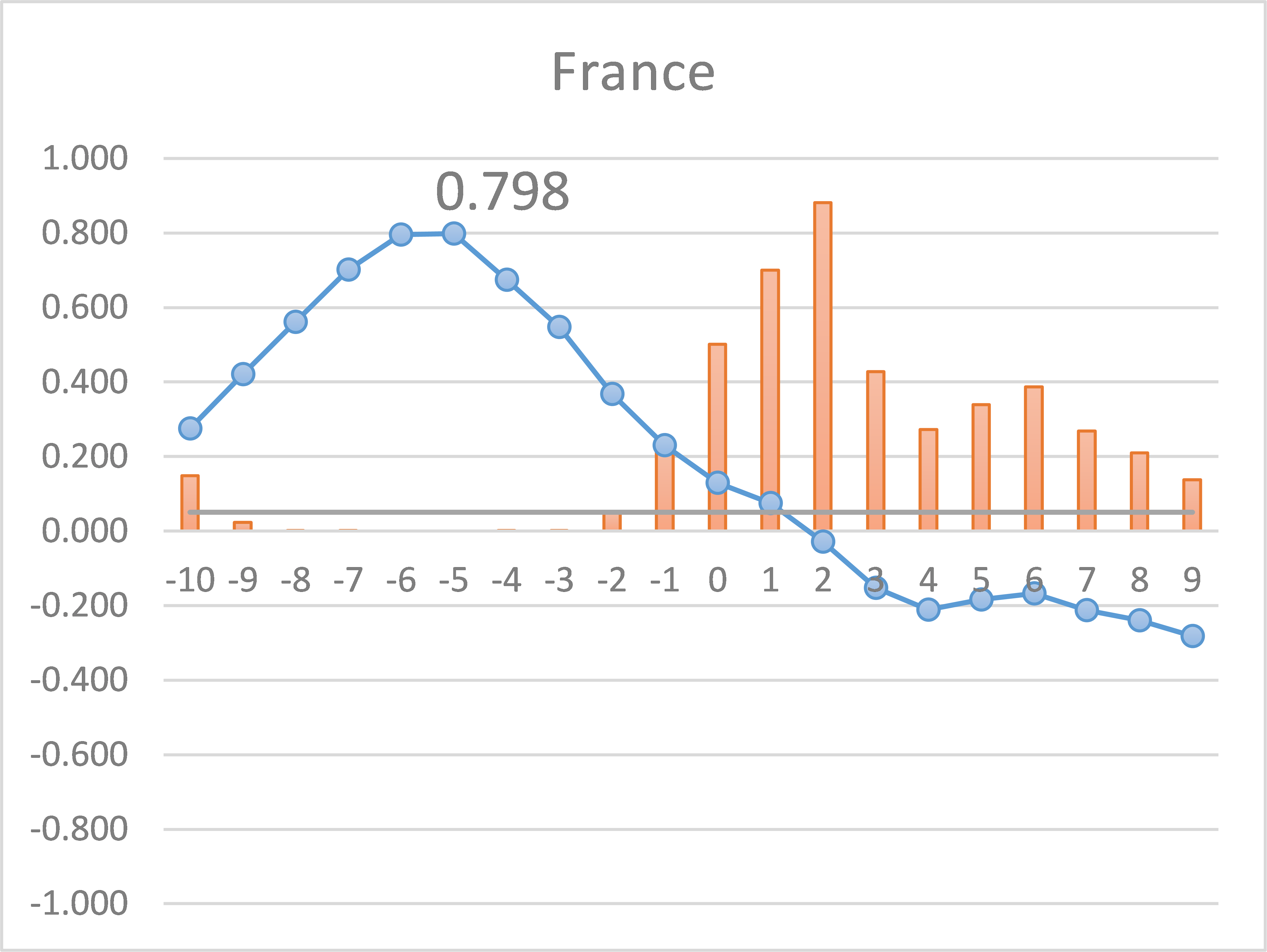}\vspace{0pt}\hspace{0pt}
\includegraphics[width=1\linewidth]{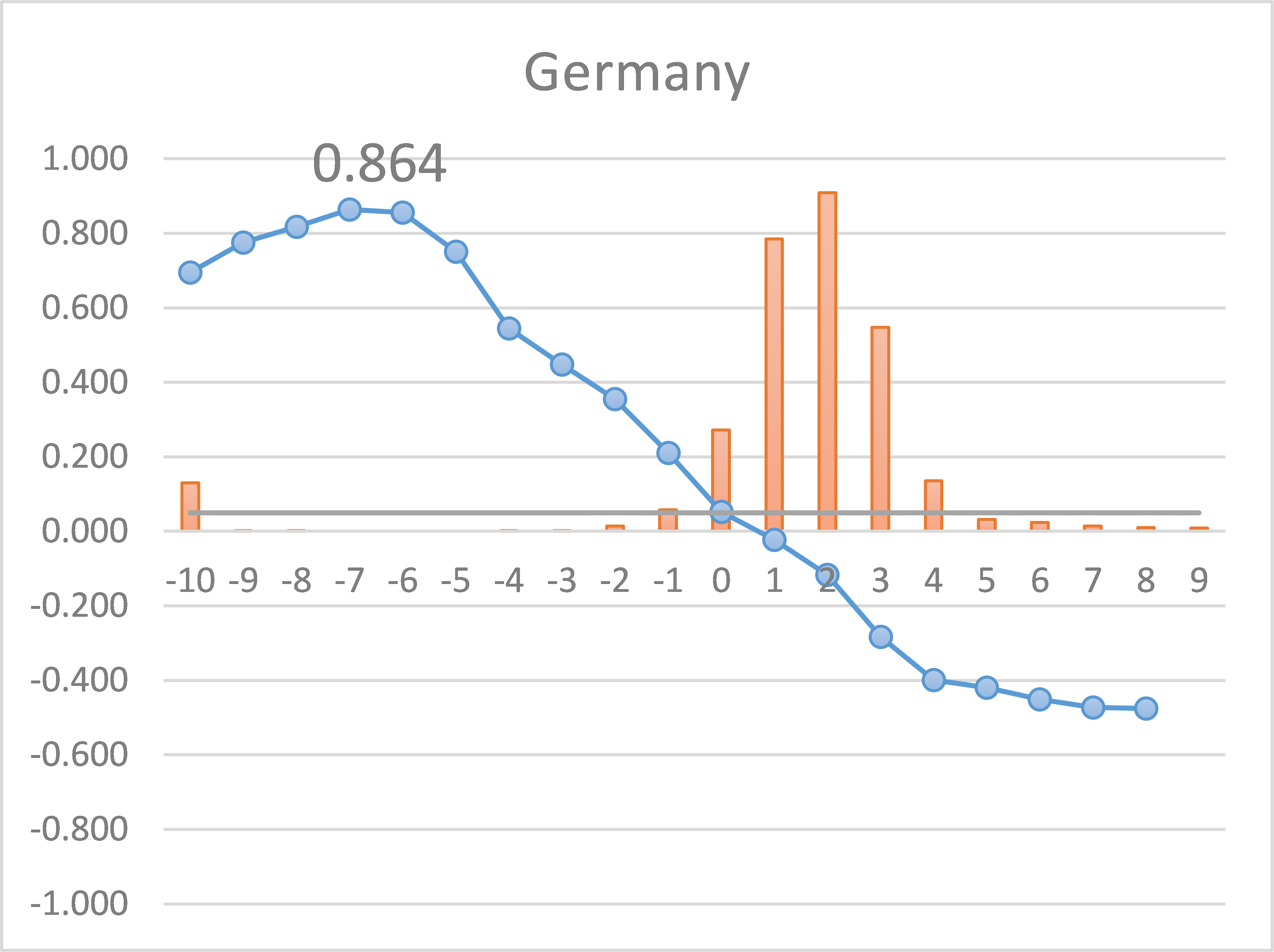}
\end{minipage}}
\subfigure[Free-contagious]{
\begin{minipage}[b]{0.23\linewidth}
\includegraphics[width=1\linewidth]{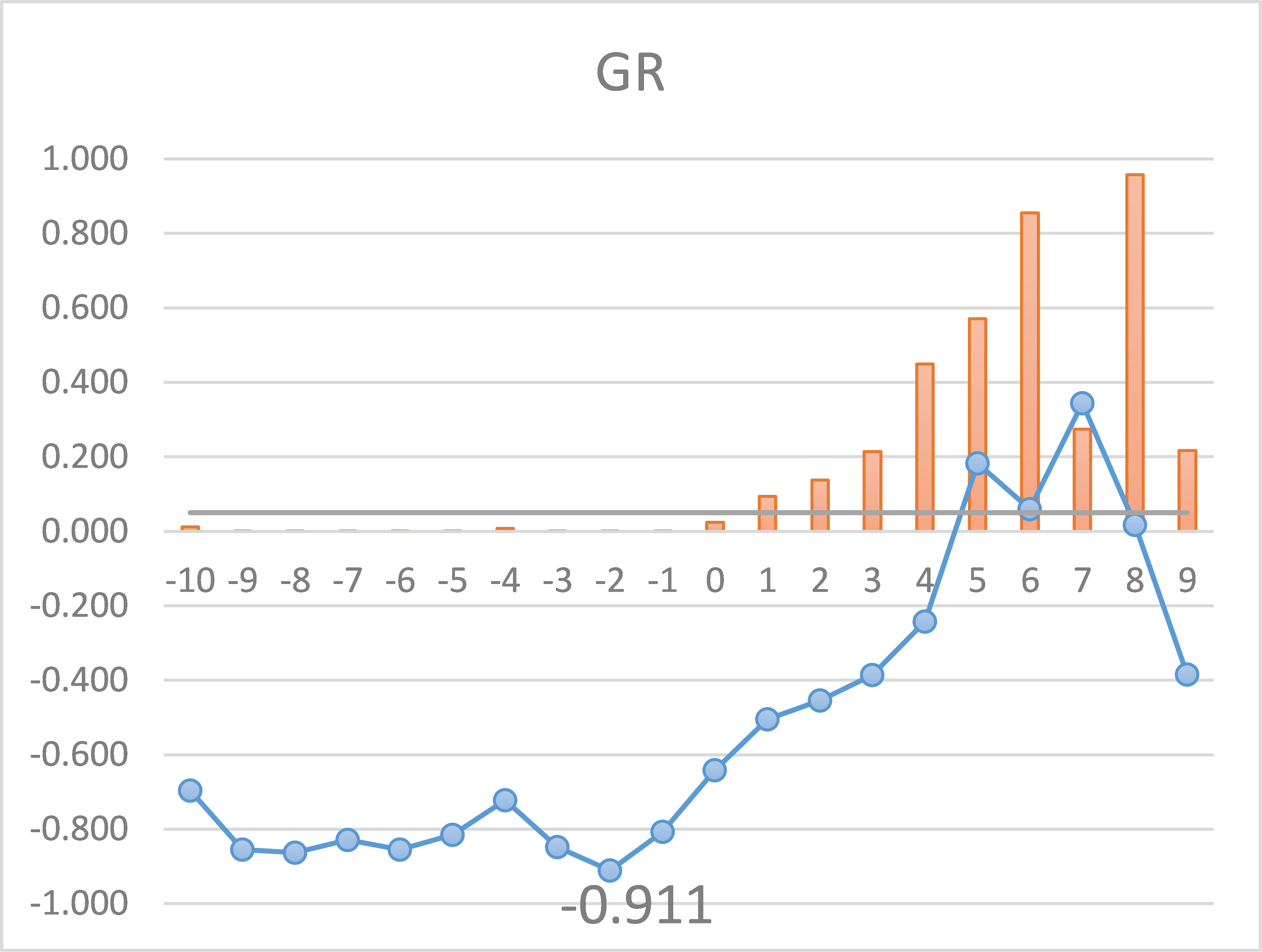}\vspace{0pt}\hspace{0pt}
\includegraphics[width=1\linewidth]{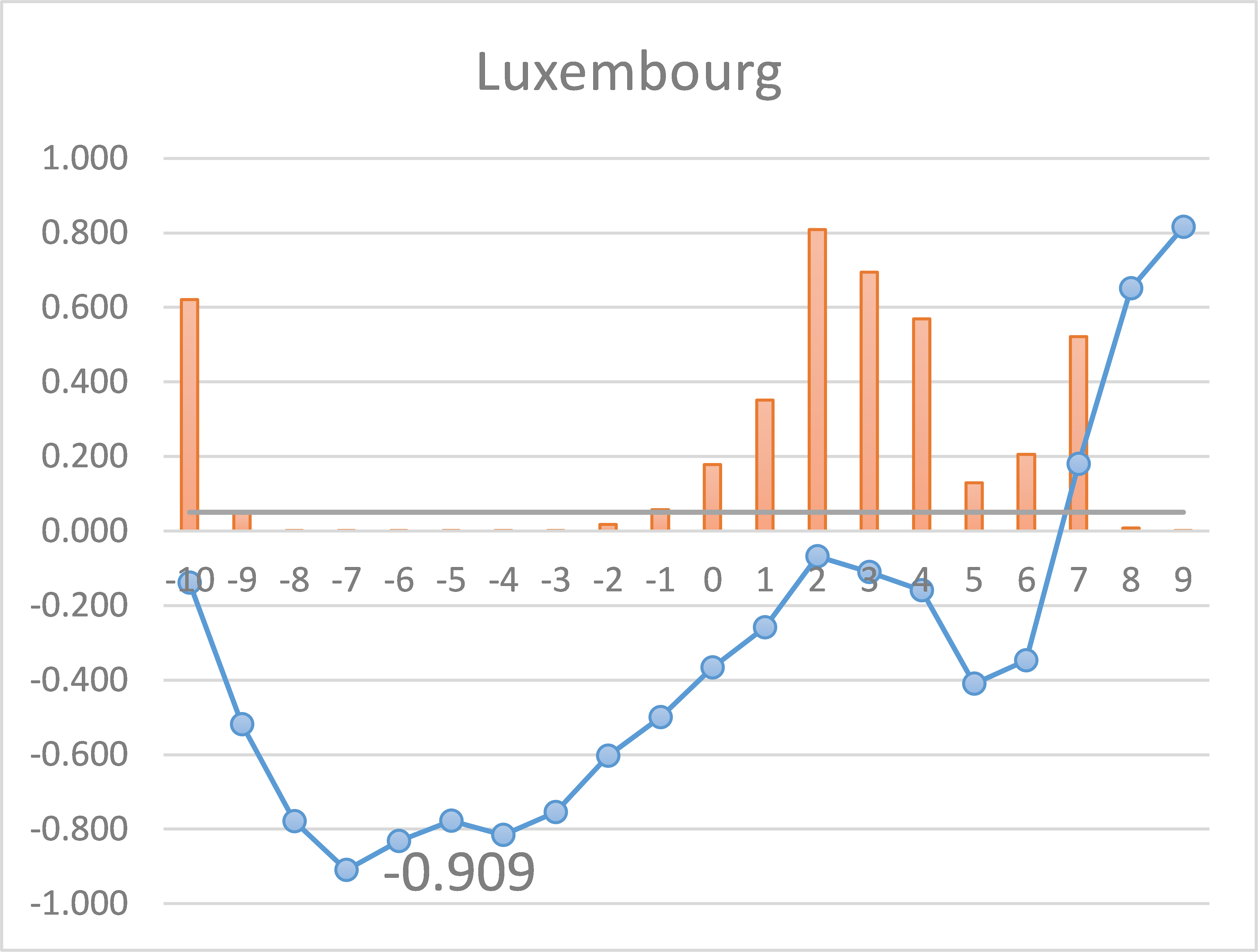}\vspace{0pt}\hspace{0pt}
\includegraphics[width=1\linewidth]{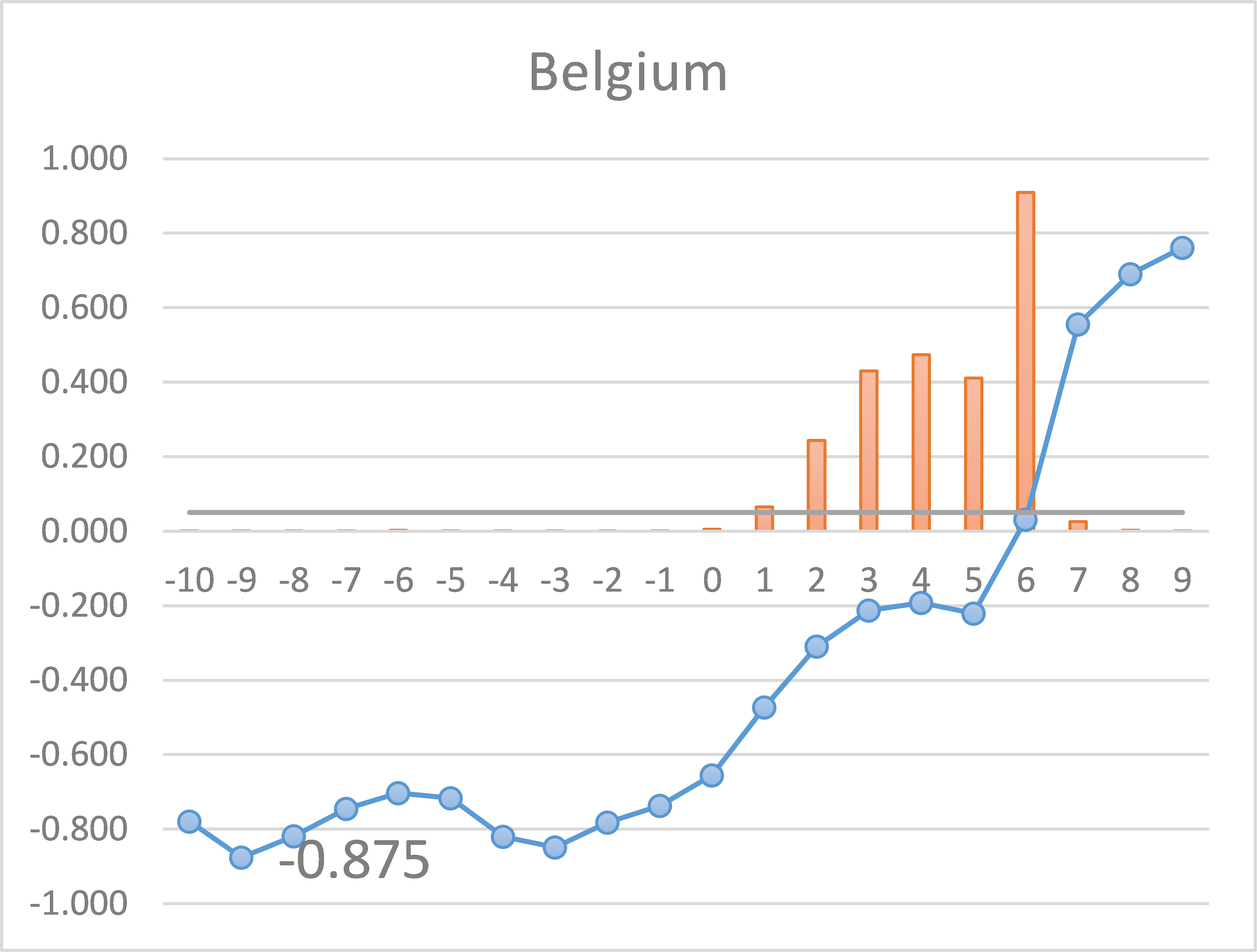}\vspace{0pt}\hspace{0pt}
\includegraphics[width=1\linewidth]{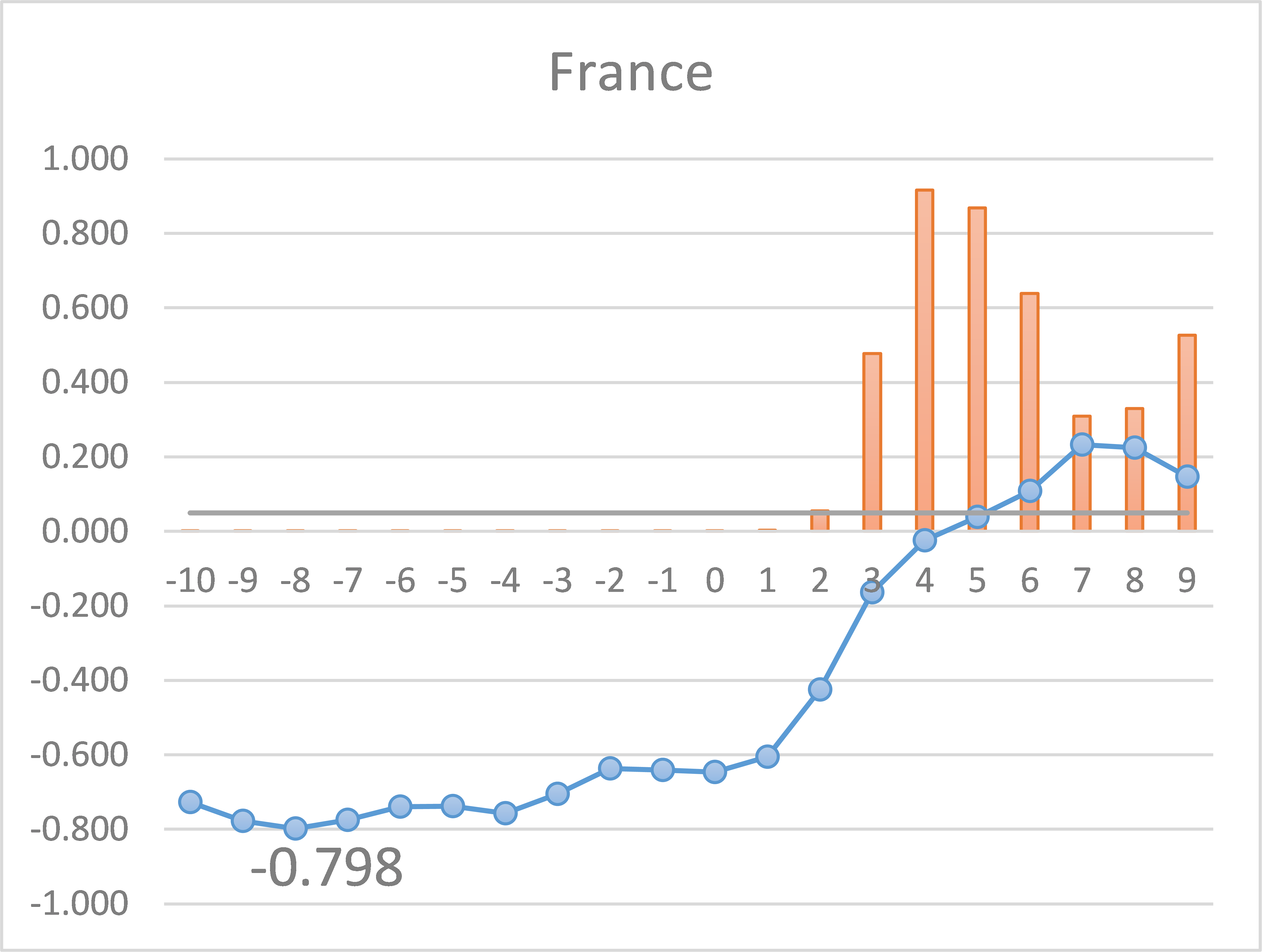}\vspace{0pt}\hspace{0pt}
\includegraphics[width=1\linewidth]{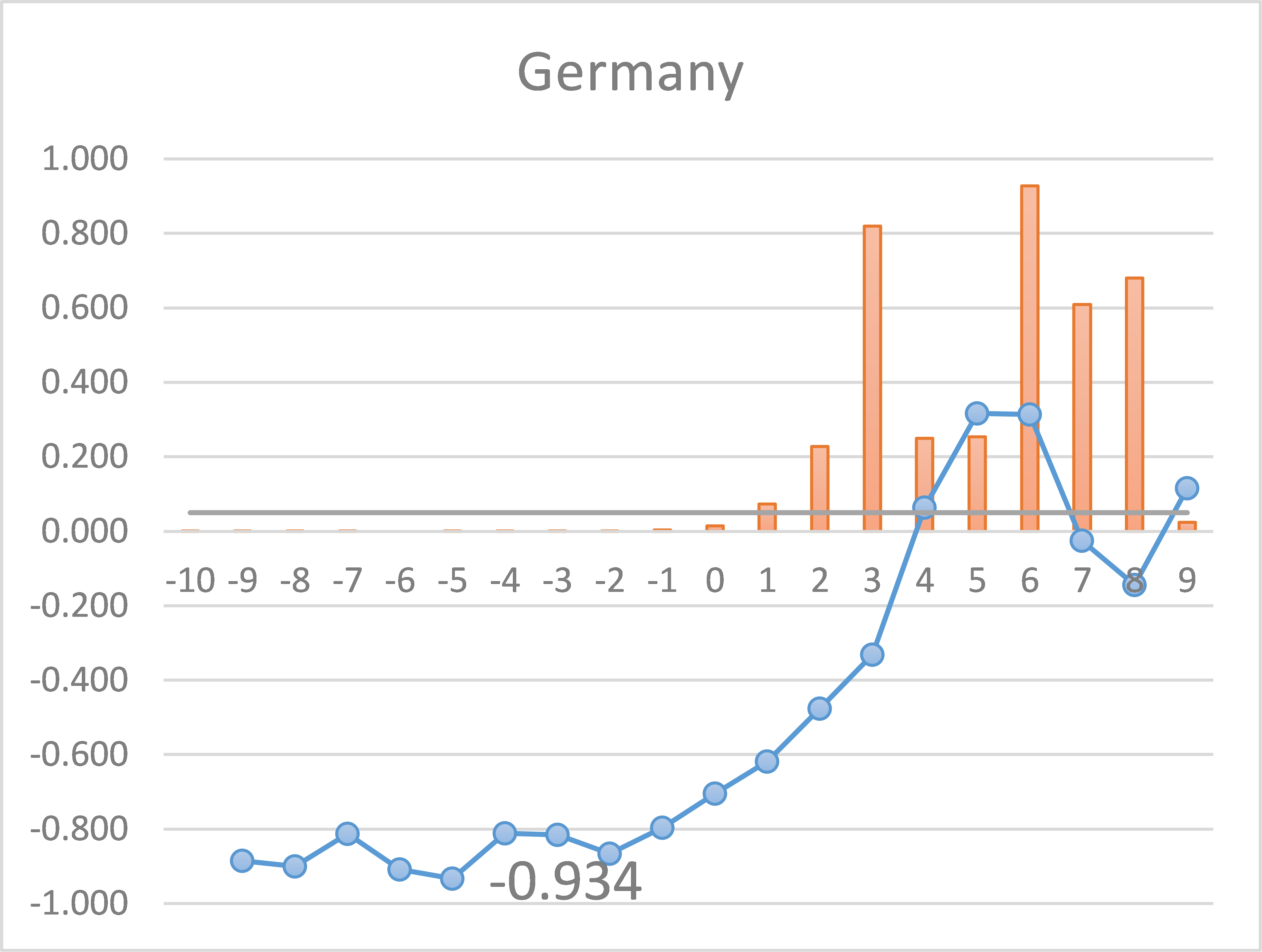}
\end{minipage}}
\subfigure[Measures]{
\begin{minipage}[b]{0.23\linewidth}
\includegraphics[width=1\linewidth]{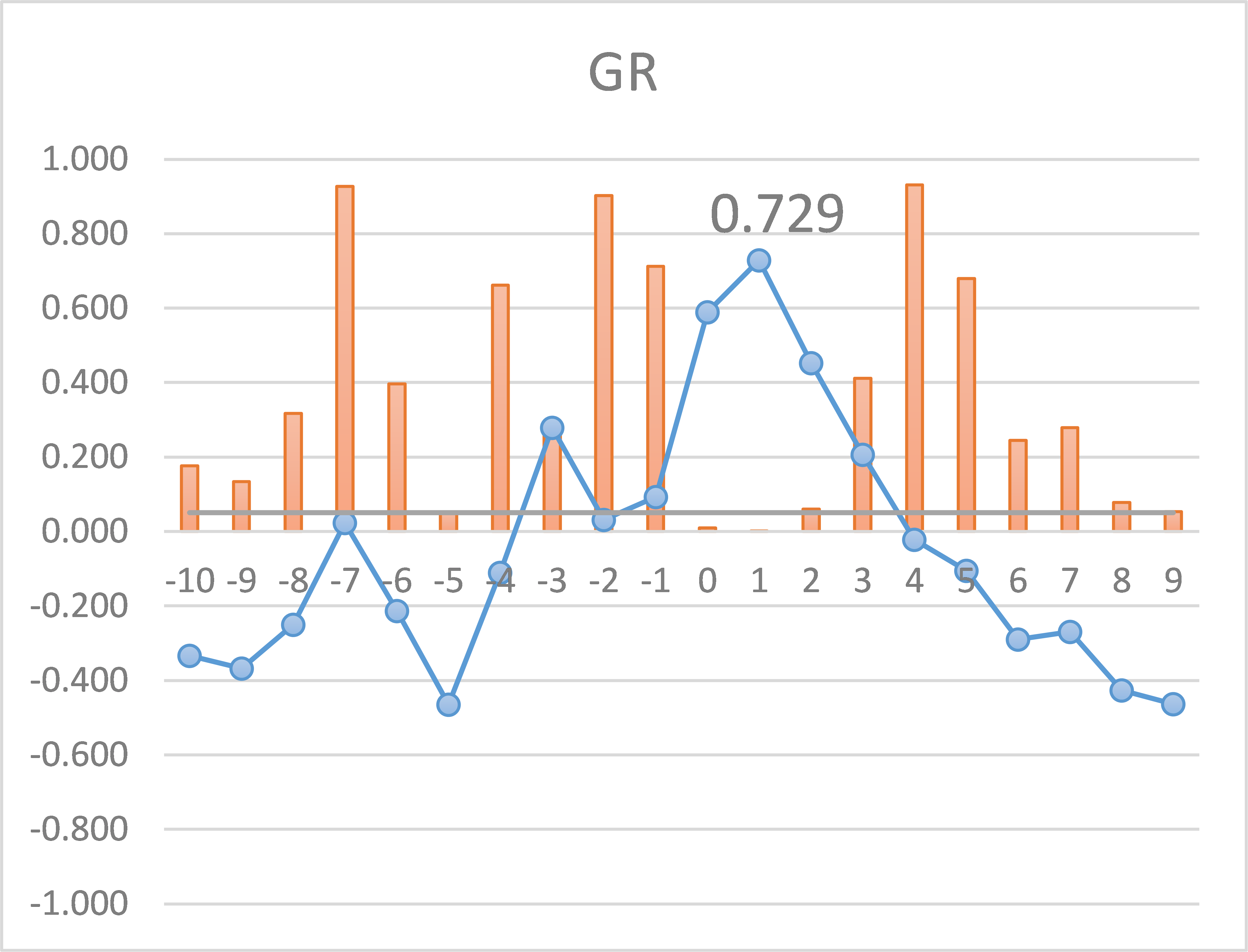}\vspace{0pt}\hspace{0pt}
\includegraphics[width=1\linewidth]{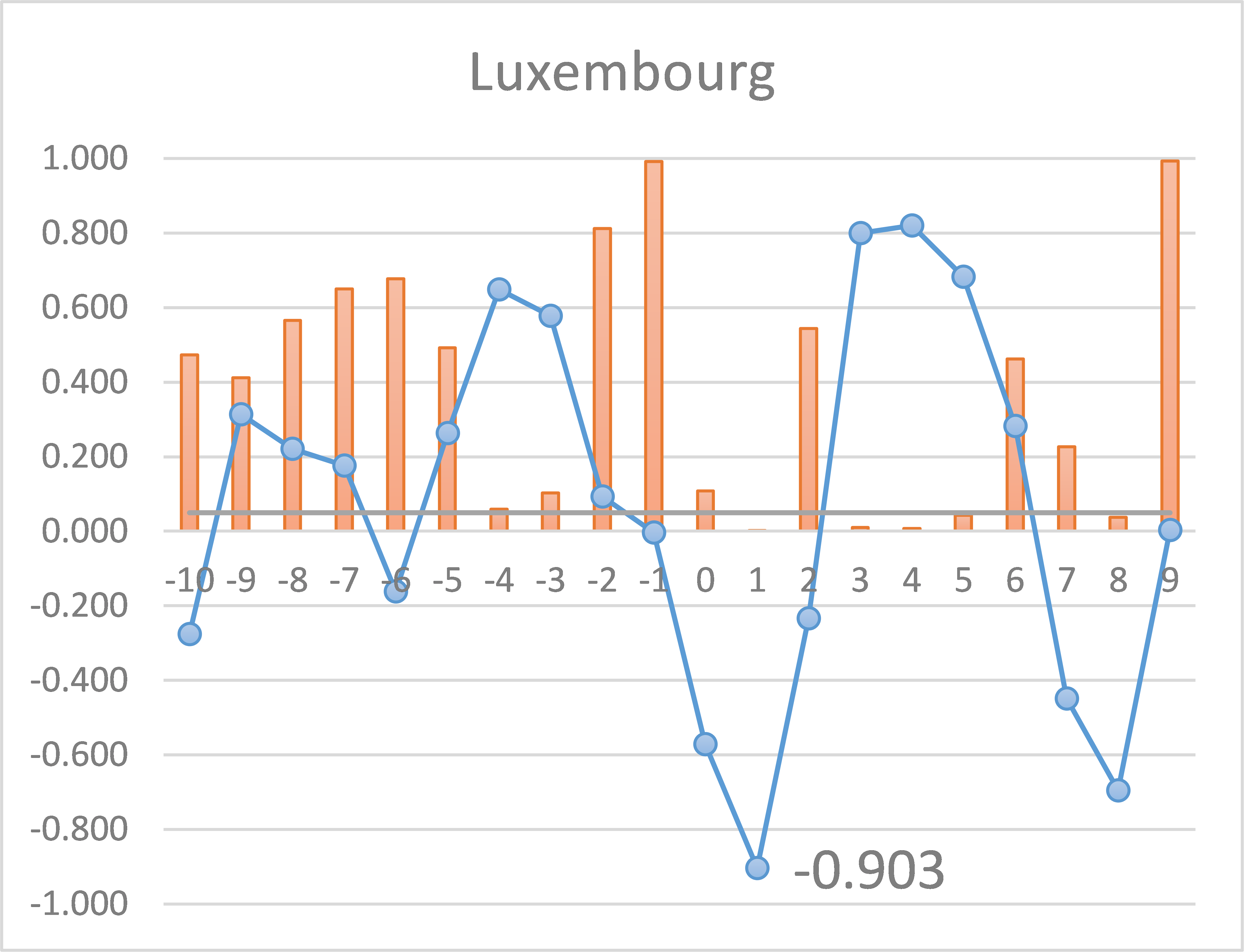}\vspace{0pt}\hspace{0pt}
\includegraphics[width=1\linewidth]{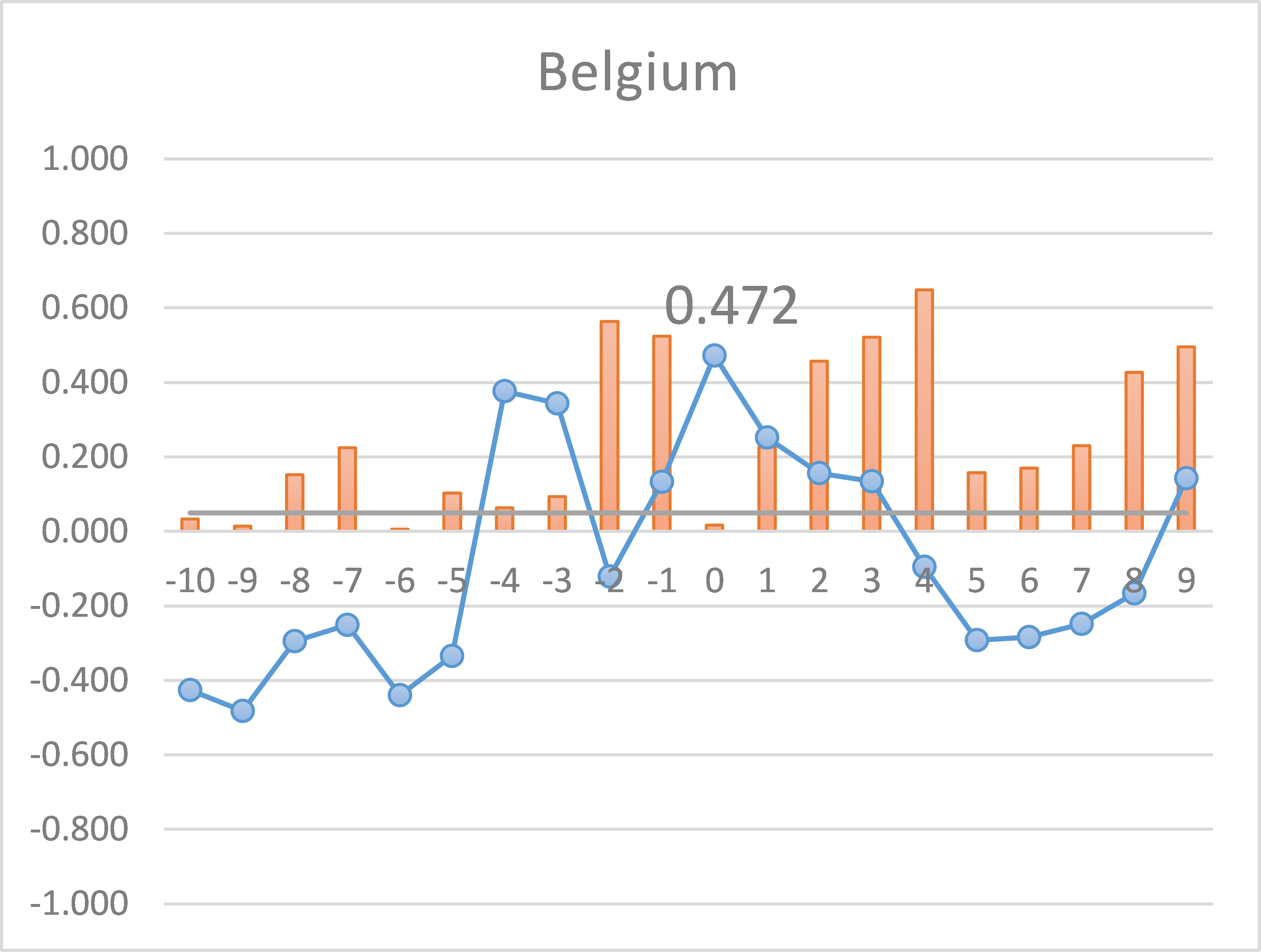}\vspace{0pt}\hspace{0pt}
\includegraphics[width=1\linewidth]{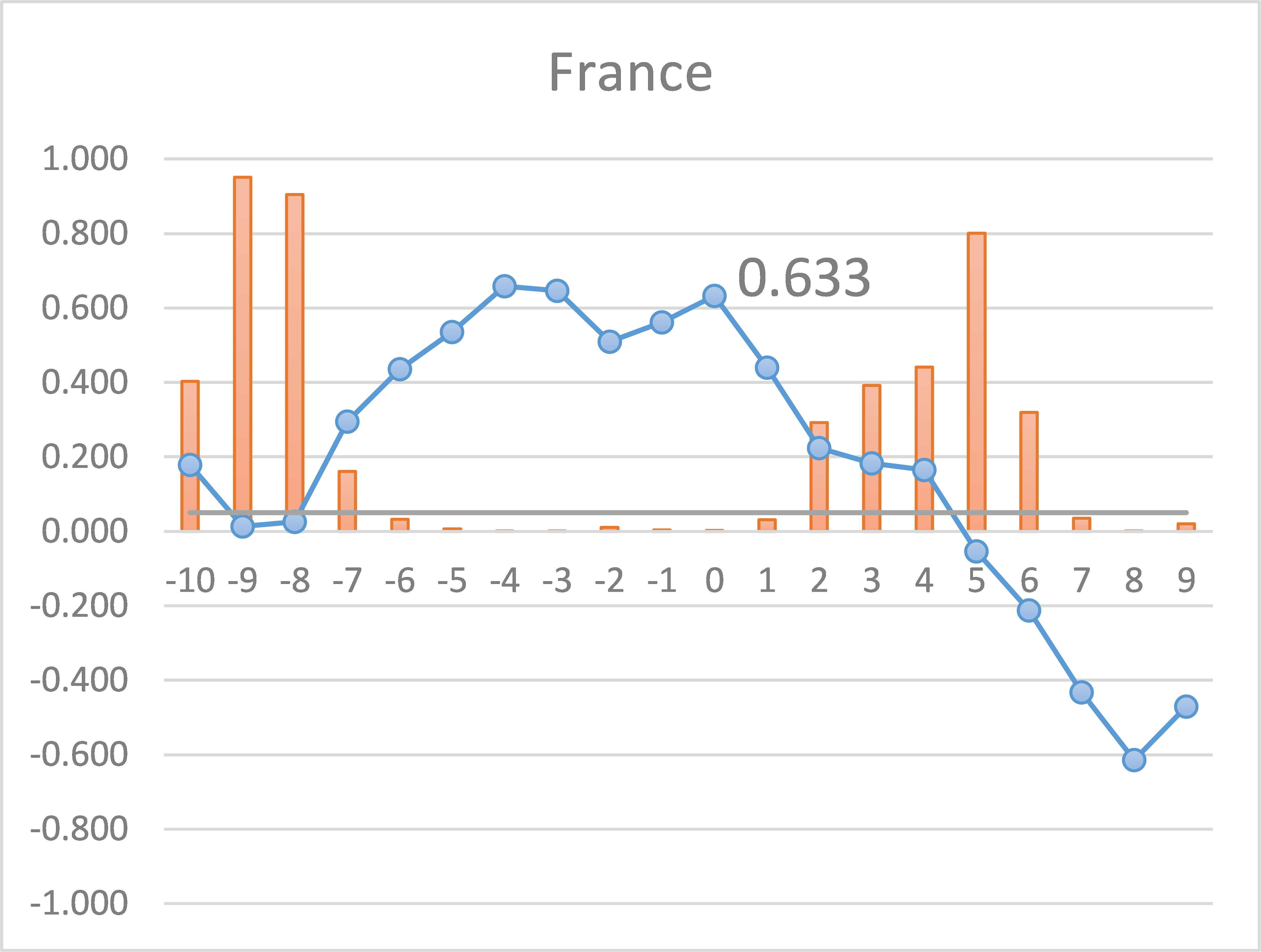}\vspace{0pt}\hspace{0pt}
\includegraphics[width=1\linewidth]{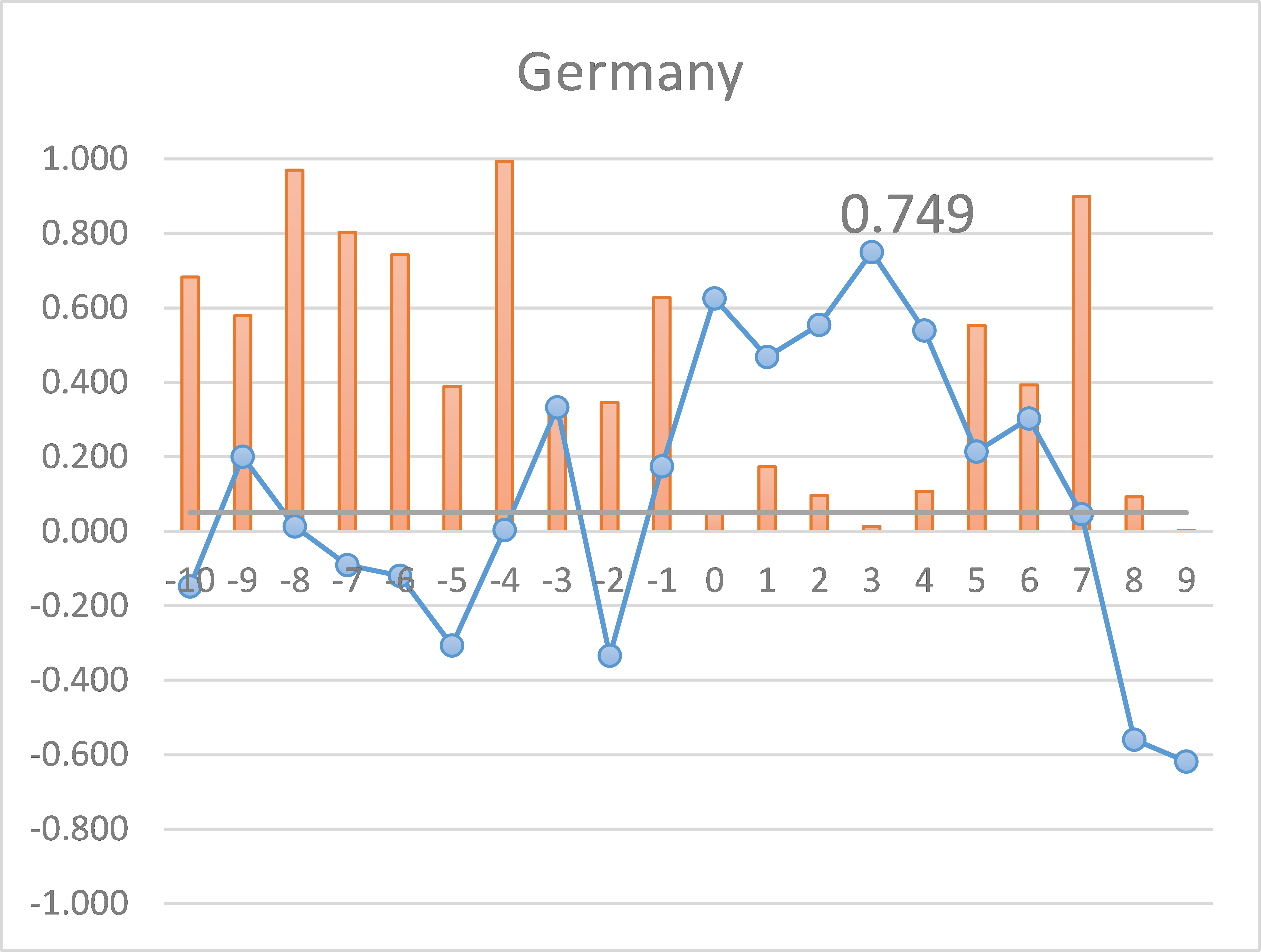}
\end{minipage}}
\subfigure[Decay]{
\begin{minipage}[b]{0.23\linewidth}
\includegraphics[width=1\linewidth]{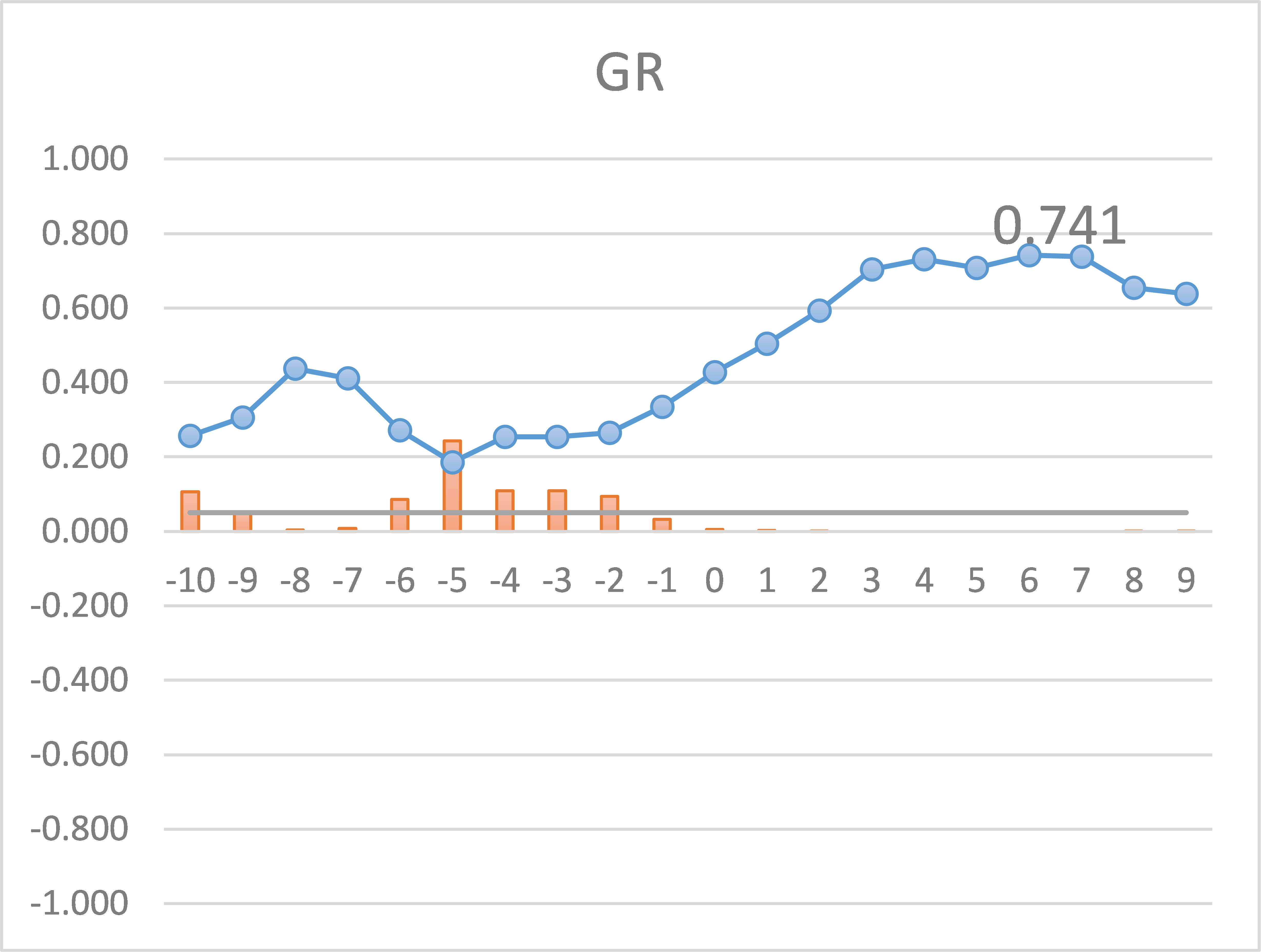}\vspace{0pt}\hspace{0pt}
\includegraphics[width=1\linewidth]{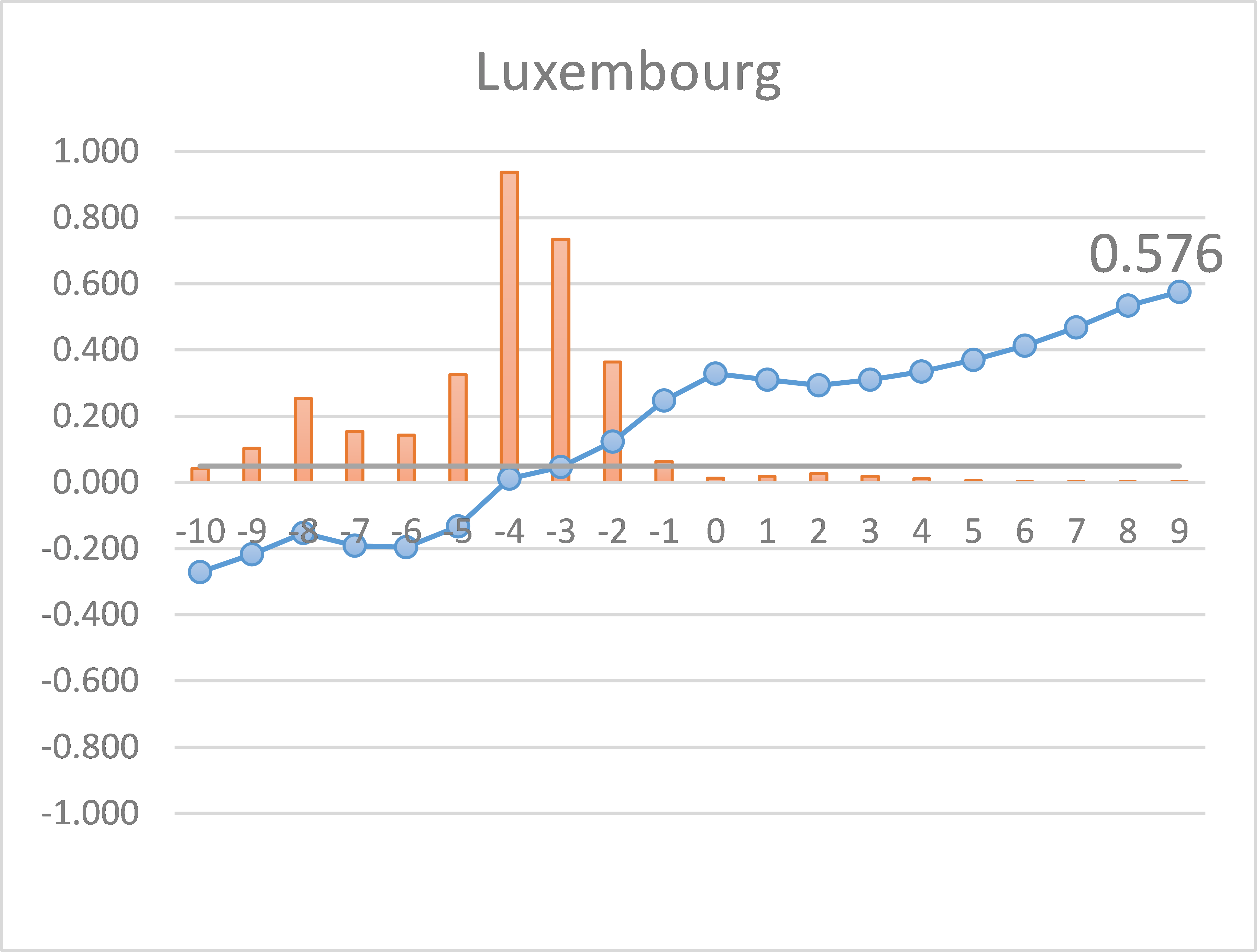}\vspace{0pt}\hspace{0pt}
\includegraphics[width=1\linewidth]{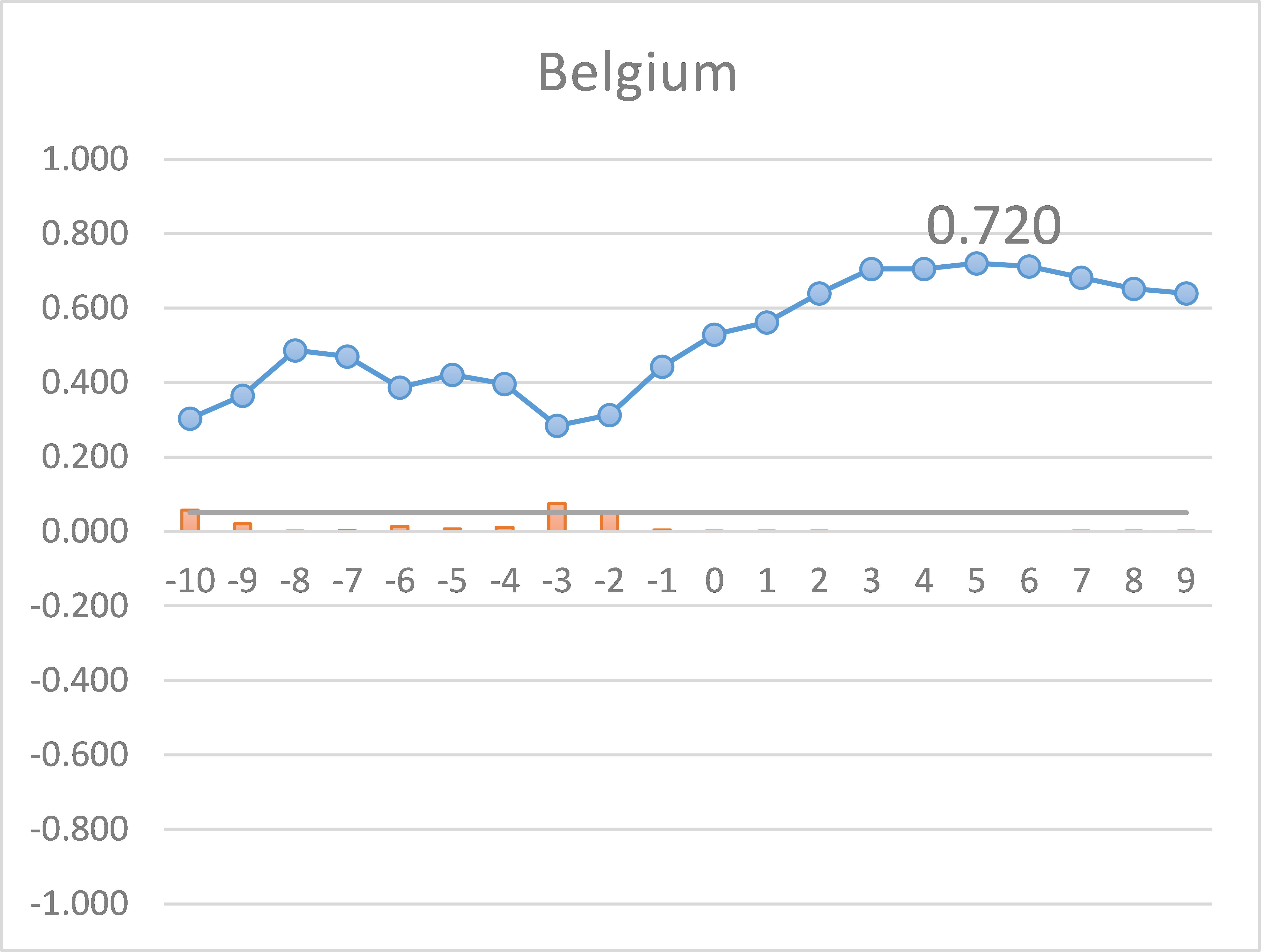}\vspace{0pt}\hspace{0pt}
\includegraphics[width=1\linewidth]{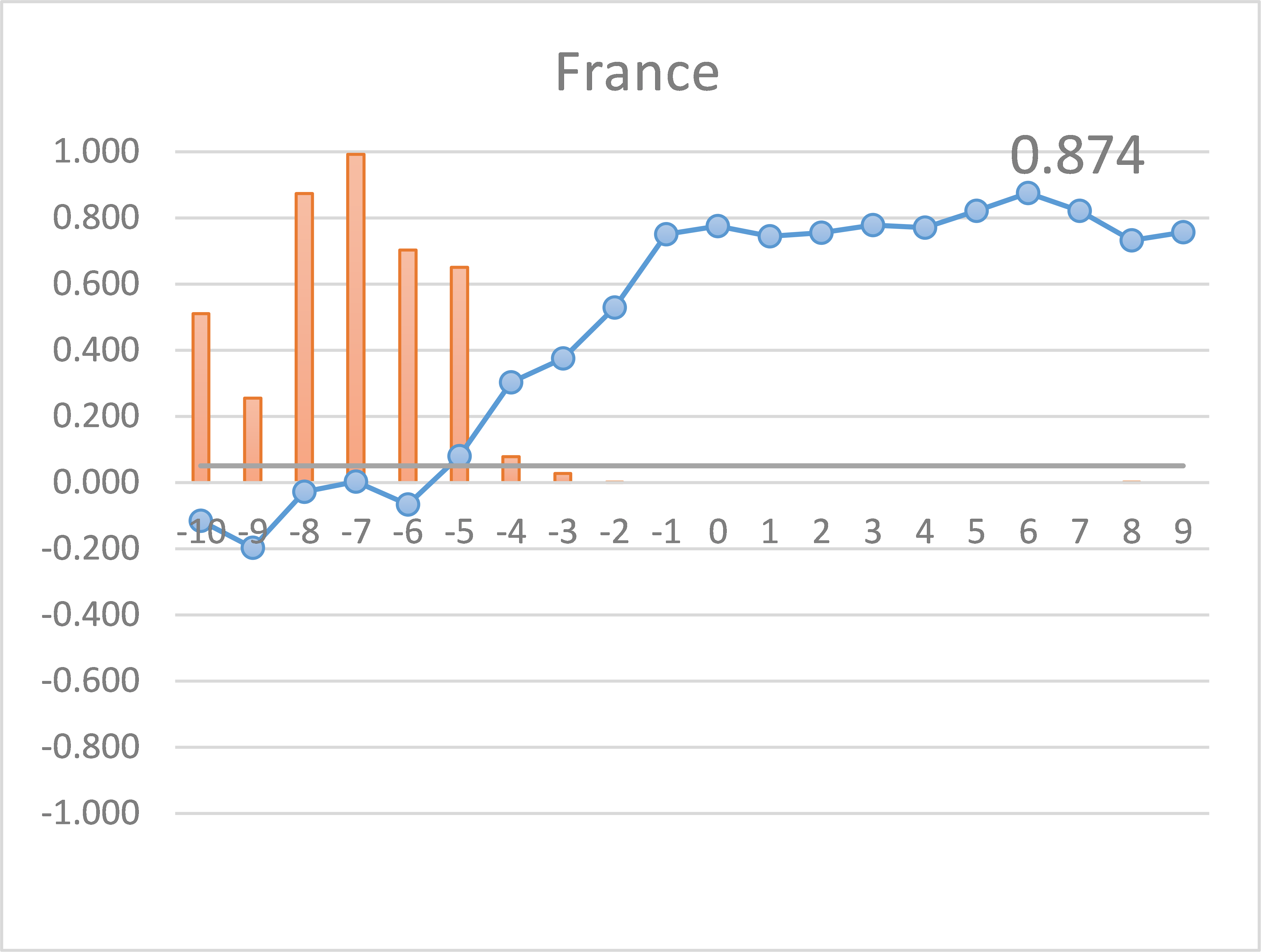}\vspace{0pt}\hspace{0pt}
\includegraphics[width=1\linewidth]{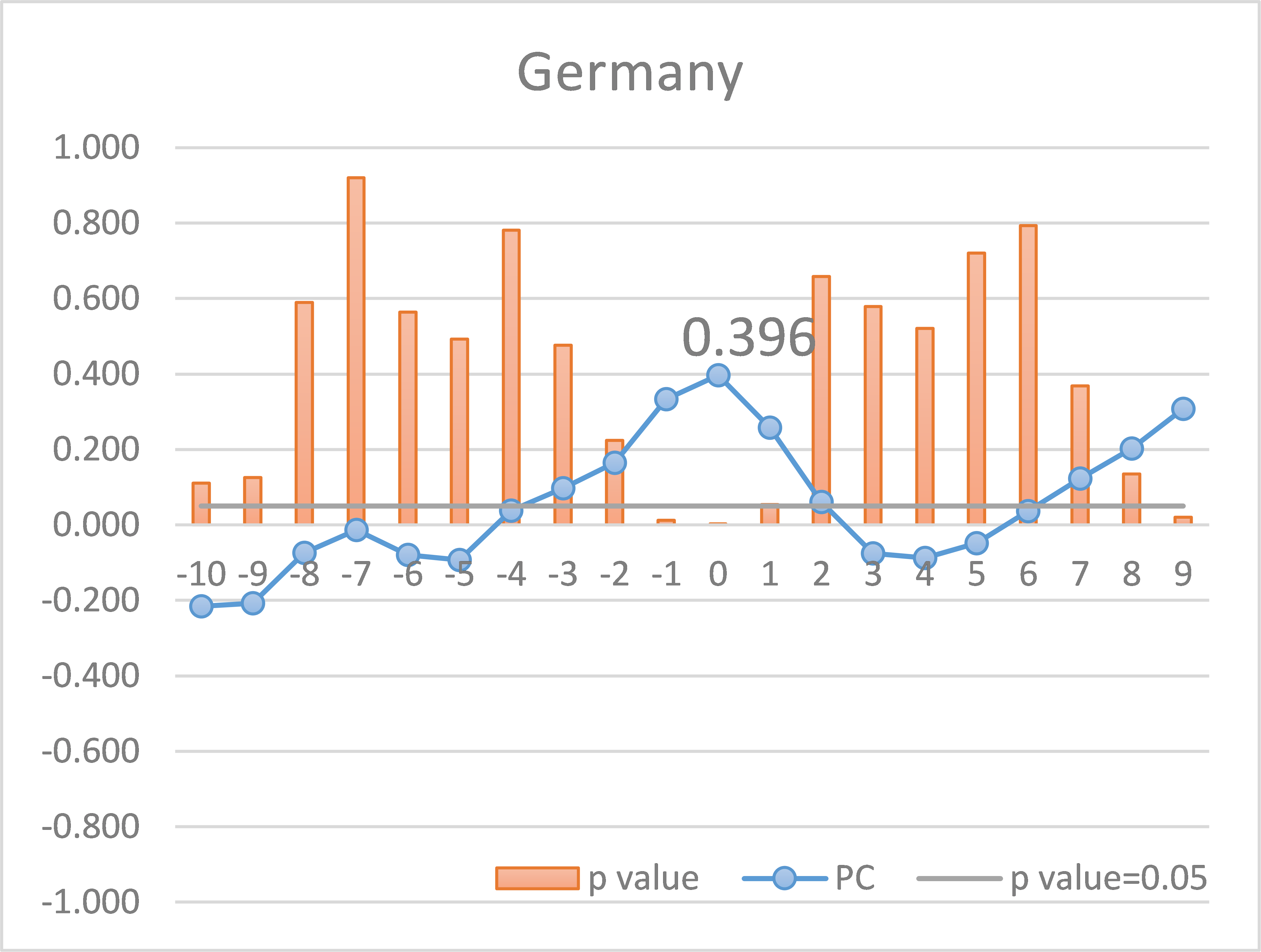}
\end{minipage}}

\caption{$PC$ (Pearson's correlation) between tweet volume and COVID-19 daily cases with different lags}
\label{fig:PPC}
\end{figure*}

We calculate the correlation between tweet volume and COVID-19 daily cases by $PC$, where a $PC$ with a large absolute value means greater relation strength. The results are shown in Figure~\ref{fig:PPC}.
A \textit{lag} refers to the tweets occurring after the cases; a Lag = -5 days means that we match the daily cases with the tweet volume from five days earlier, in other words, a 5-days lead. 

\smallskip\noindent
\textbf{{\it Pre-peak period}.} As shown from Table~\ref{fig:PPC}, there is a clear trend of strong correlation ($PC>0.8$, $p<0.05$) with lags during the {\it Pre-peak} period, reaching its' maximum at -5 or -6 days, indicating that a correlation exists between tweet volume and COVID-19 daily cases and tweet volume can help predict COVID-19 daily cases in this period.
This is highly consistent to the conclusions presented in the existing studies~\cite{singh2020first, st2012can,jahanbin2020using,younis2020social}.

\smallskip\noindent
\textbf{{\it Free-contagious period}.} There is no clear trend of correlation with lags except the value of Luxembourg, indicating that tweet volume cannot help predict the daily cases in the {\it Free-contagious} period. The period only lasted for 4 days in Luxembourg, which is too small to make $PC$ a reflection of the correlation. However, the $PC$ values show a highly negative correlation between tweet volume and daily cases. This indicates that there is a short downward trend in the discussion of the pandemic after it reached its peak, even though the number of cases continued to rise rapidly. This result validates the conclusion of Smith et al.~\cite{smith2016towards} from our dataset, who noted that public concerns of disease decline sharply after the peak even though the infection rates remain high. In other words, the public concerns of the pandemic decline after the {\it Pre-peak} period. 

\smallskip\noindent
\textbf{{\it Measures period}.} There is a clear trend of correlation with lags,  tweet volume begins to level off, with a 0 or 1-day-lag moderate correlation ($0.8>PC>0.3$, $p<0.05$) to the daily cases. Tweet volume cannot help predict daily cases here because it fluctuates with the number of cases on the current or previous day. It is worth noting that Pearson's coefficient is sensitive to outliers and is not robust. With too few dates included, a single outlier can change the direction of the coefficients. This period existed for only 8 days in Luxembourg, resulting in an anomaly value ($PC = -0.903$). 
It is assumed here that fluctuating changes in tweet volume during this period are influenced by local news and policies, and further discussion will take place in Section~\ref{sec:topics}.

\smallskip\noindent
\textbf{{\it Decay period}.} The correlations between tweet volume and daily cases occur in two ways. One is weakly correlated, the other reveals a correlation, but the trend of correlation with lags is insignificant. Both ways demonstrate that it is not possible to estimate daily cases with the help of tweet volume during this period.

In summary, with the Spatio-temporal analysis of the correlation between tweet volume and COVID-19 daily cases during the four periods of the pandemic, we reject the hypothesis that there is a strong correlation between tweet volume and COVID-19 daily cases in GR and related countries ({\sf H1}) and tweet volume can help predict COVID-19 daily cases ({\sf H2}). More accurately, {\sf H1} and {\sf H2} can only be confirmed during the {\it Pre-peak} period. In this period, regardless of the time at which $R(t)$ peaks, there is a 5-6 day lead between tweet volume and COVID-19 daily cases.
Moreover, before the pandemic strikes, there is a high level of tweet volume regarding the pandemic. On the particularity of GR, we find that the {\it Free-contagious} period in GR and Luxembourg are exceedingly shorter (6 and 4 days, respectively), during the {\it Measures} period.

\section{Topic Modelling and Classification of Tweets}
\label{sec:topics}

\begin{figure}[!t]
\centering
\includegraphics[width =0.95\textwidth]{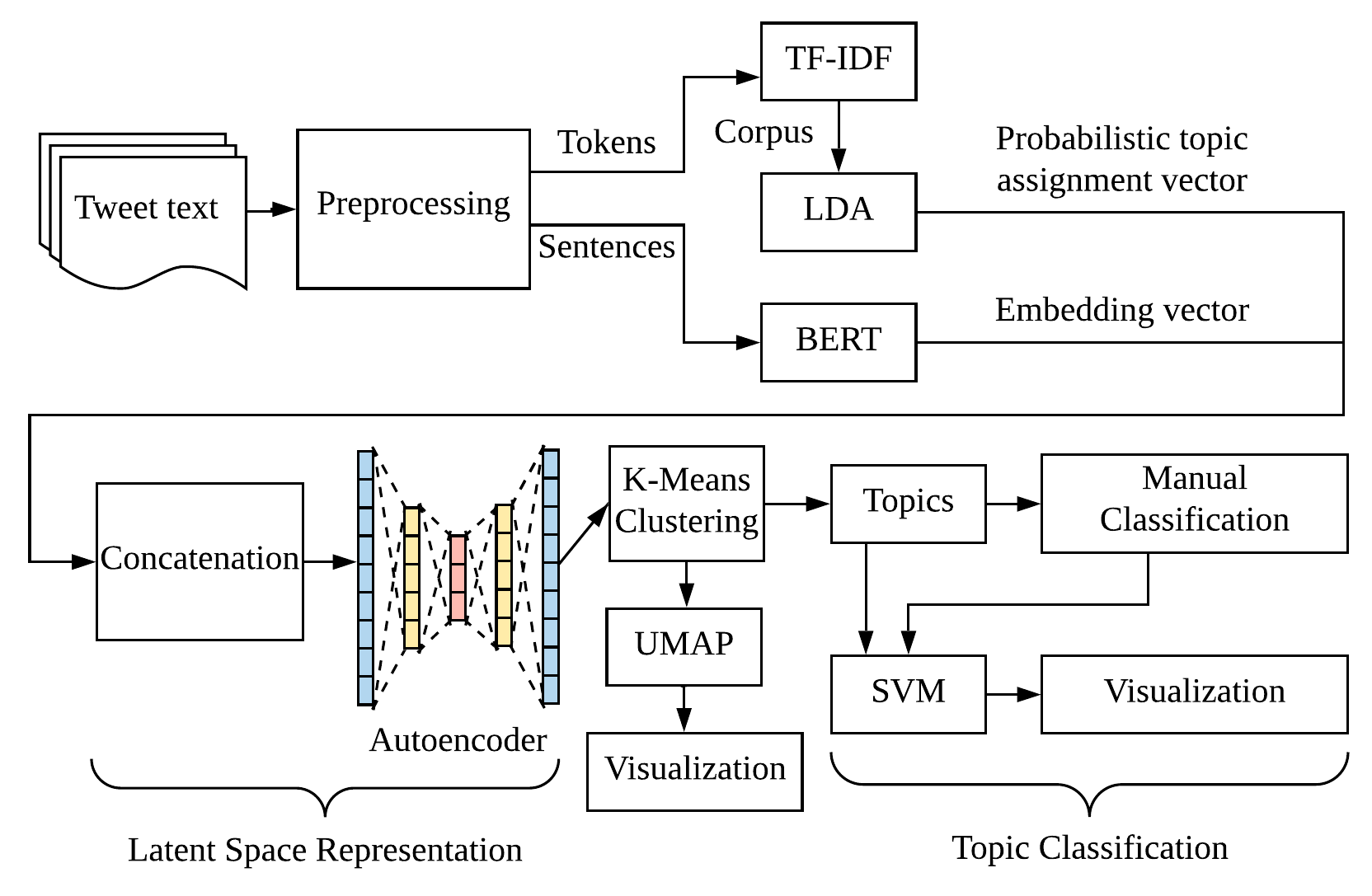}
\caption{Workflow of topic modelling and classification}
\label{fig:pipeline}
\end{figure}

In the previous section, we conduct an overarching preliminary analysis of tweet volume, but without the in-depth discussion of tweet text.
In this section, we build a workflow to analyse tweet text as shown in Figure~\ref{fig:pipeline}. This workflow includes tweet text pre-possessing, topical modelling, and classification of the generated topics, each part is described in details below.
We perform topical modelling on the tweet text to extract the main topics discussed every day in each region and country, and then train a classifier to distinguish these topics into 7 categories in order to observe and analyse the changes in the topics discussed in each region and country during different periods of the pandemic. In parallel, we observe and investigate whether there are distinctive characteristic of these region and countries’ topics about COVID-19 on Twitter, and focus on the differences that exist in the GR.

\subsection{Text prepossessing and topic modelling}
\smallskip\noindent
\textit{Text prepossessing.} Prior to topical modelling, the tweets data needs to be preprocessed. All text are lower-cased, while URLs that mention  usernames and `RT' are removed as well.
Besides, punctuation and numbers are filtered out, typos are corrected by Symspell\footnote{\url{https://github.com/wolfgarbe/symspell}}
and stop words are removed.
Then since the tweets are collected based on the keyword search, we removed keywords such as `coronavirus', `koronavirus', `corona', `covid-19', `covid' from the text to avoid bias. 
Finally, Natural Language Toolkit (NLTK) is used for tagging the part-of-speech, stemming and tokenization. Nouns, verbs, adjectives and adverbs are selected.

\smallskip\noindent
\textit{Topic modelling.} Aiming to identify the latent topics of the tweets  posted by the public in GR and related countries, we adopt the general structure of contextual topic embedding method (CTE)\footnote{https://bit.ly/3hUQjzf} to extract daily topics and get a more accurate picture of topic trends.
CTE mainly consists of two components, LDA and BERT, to extract different information from sentences to embedding.
LDA, a bag-of-words approach which is widely used to identify latent subject information in a large-scale archives or corpus has its drawback: it needs large corpus to train, ignores contextual information and performs mediocrely in handling short texts~\cite{yan2013biterm}. 

BERT utilises bidirectional transformers for pre-training on a large unlabelled text corpus, taking both left and right context into account simultaneously, which compensates for the shortcoming of LDA.
Also, BERT is a method available for sentence embedding, thus we concatenate the generated tokens of each tweet text as input sentences for BERT to obtain sentence embedding vectors.
CTE combines the sentence embedding vector generated by BERT with the probabilistic topic assignment vector generated by LDA with a hyper-parameter $\gamma$. 
After obtaining the concatenated vector in high-dimensional space, CTE uses an autoencoder to learn a low-dimensional latent space representation of the concatenated vector with more condensed information.
Then $k$-means~\cite{wagstaff2001constrained} is implemented for clustering, and the number of clusters $k$, that is, the number of topics, reserved as a hyper-parameter.
We extract the word frequency in each cluster, sort and then take the top ten as the representative topics of that cluster.
In terms of visualisation, Uniform Manifold Approximation and Projection (UMAP)~\cite{mcinnes2018umap} is used for low-dimensional latent space degradation, which is the state-of-the-art visualisation and dimension reduction algorithm.

The CTE rather than a single LDA model is chosen as our topical modelling approach due to the fact that LDA is designed for monolingual contents and lacks the structure necessary to generate effective multilingual topics~\cite{gutierrez2016detecting}.
GR, as a relational city, are multilingualism. 
CTE includes BERT, a sentence embedding model that can handle multi-language, can tackle this problem.
Two adjustments are therefore made to the original CTE. For one, we adopt the BERT-based multilingual model as the pre-trained model in BERT\footnote{https://github.com/google-research/bert/blob/master/multilingual.md}.
In addition, some words appear less frequent than in English which is predominantly spoken and are easily overlooked in LDA. Thus, we adopt the TF-IDF model to determine word relevance in the documents~\cite{ramos2003using}. We further feed the generated corpus by TF-IDF to LDA, instead of simple bag-of-words corpus.

\begin{table}[!h]
\centering
\begin{tabular}{|l|r|r|}
\hline
Country    & \multicolumn{1}{l|}{Coherence score} & \multicolumn{1}{l|}{Silhouette score} \\ \hline
GR         & 0.432                                & 0.893                                 \\ \hline
Luxembourg & 0.474                                & 0.894                                 \\ \hline
France     & 0.351                                & 0.590                                 \\ \hline
Belgium    & 0.377                                & 0.864                                 \\ \hline
Germany    & 0.336                                & 0.655                                 \\ \hline
\end{tabular}
\centering
\caption{Average coherence score and average silhouette score of CTE}
\label{CS&SS}
\end{table}

Average coherence score~\cite{o2015analysis,newman2010automatic} and average silhouette score~\cite{aranganayagi2007clustering} are utilised as the metrics of CTE. We calculated an average coherence score by calculating the topic coherence for each topic individually and averaging them. The hyper-parameters are tuned to obtain the best results. The value of $k$ is chosen from $\{1,2,\dots,15\}$ and the value of $\gamma$ is chosen from $\{0.1,0.2,\dots,0.9\}$. The model arrive at the optimal with $k = 7$ and $\gamma = 0.5$.

\begin{figure}[!t]
\centering
\includegraphics[width =0.95\textwidth]{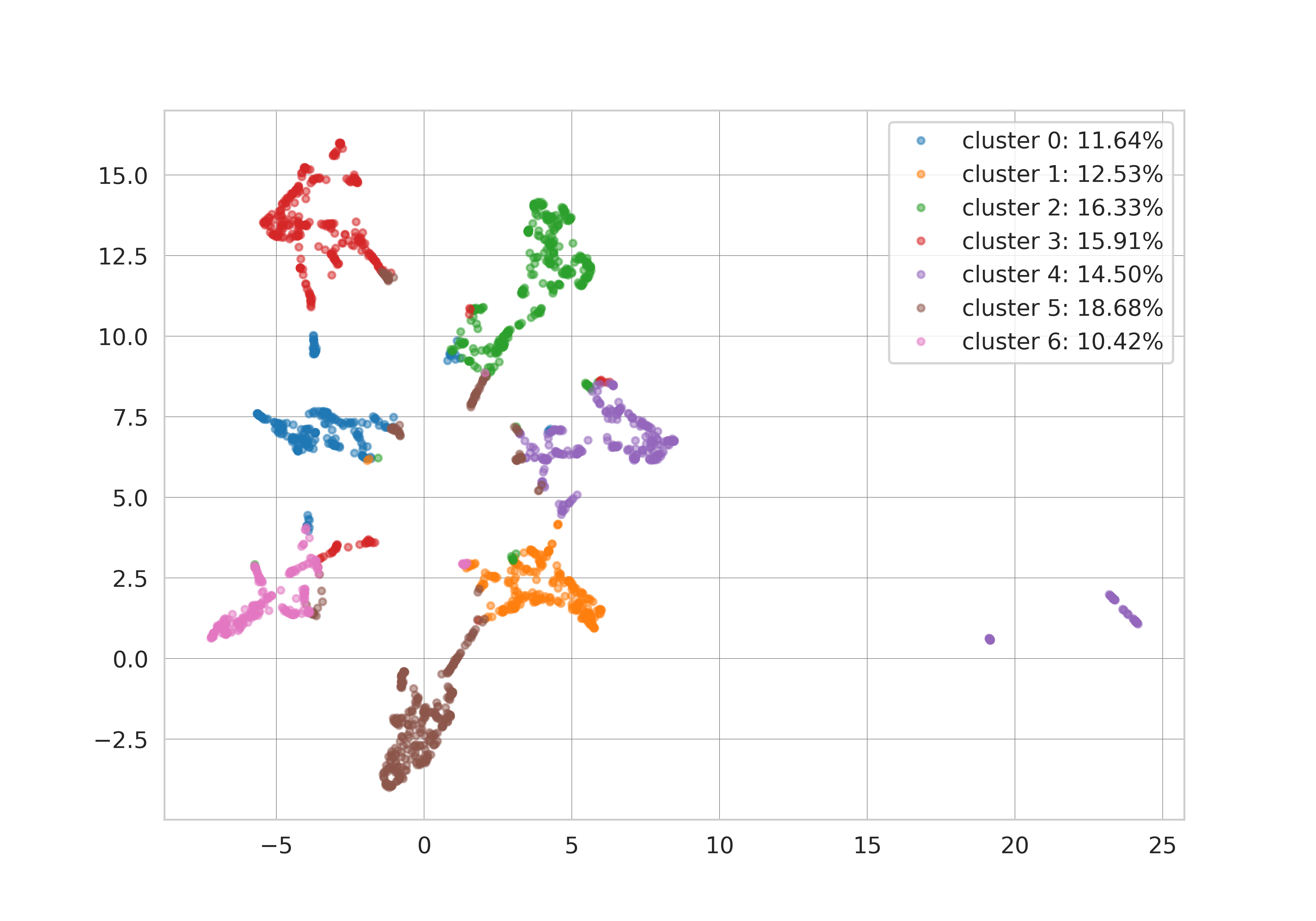}
\vspace{-5mm}
\caption{A sample of UMAP clustering results}
\label{fig:UMAP_sample}
\end{figure}

The results are shown in Table~\ref{CS&SS} and a sample of clustering result from UMAP is shown in Figure~\ref{fig:UMAP_sample}. It can be observed from Table~\ref{CS&SS} and Figure~\ref{fig:UMAP_sample} that the results generated by CTE are coherent and can be observed as well-separated clusters.

\subsection{Topic classification}
After getting 4,763 topics from topic modelling, we then randomly selected 2,435 topics and classified manually into the following 7 categories:

\begin{enumerate}
\item`Wuhan \& China': Topics about Wuhan and China.
\item`Measures': Topics about basic information including symptoms, anti-contagion and treatment measures of COVID-19.
\item `Local news': Topics about local COVID-19 news, including daily new cases, deaths, etc.
\item`International news': Topics about international COVID-19 news  
\item `Policy and daily life': Topics about COVID-19 related policies encompass lockdown, closure of borders, limits on public gatherings and the impact of the policies on daily life.
\item `Racism': Topics about racism.
\item `Other': Other topics.
\end{enumerate}

The division of these 7 categories is based on the classification of COVID-19 related Twitter topics analysis in existing studies \cite{abd2020top,ordun2020exploratory}, and is determined empirically on the basis of common knowledge and the status quo.

These manually classified topics are used to train a Support Vector Machine (SVM)~\cite{chang2011libsvm} for supervised classification. The reasons for training a classifier instead of manually labelling all the topics are, on the one hand, the classifion of all the topics manually is time-consuming, and, on the other hand, the classifier can be used in further studies.

Words of each topic are converted to word frequency vectors with TfidfVectorizer\footnote{\url{https://bit.ly/30bA8Ye}} and country are encoded with Label Encoder\footnote{\url{https://bit.ly/39EO5kK}}. The feature vector is consisted by these two elements.
Since our manually labelled dataset is imbalance in classification, Synthetic Minority Oversampling Technique~\cite{chawla2002smote} is utilised for oversampling imbalanced the dataset and mitigate imbalances. 
The dataset is split, $80\%$ of which is the training dataset and $20\%$ the test dataset. Grid search with 10-fold cross-validation is deployed on training dataset to find the optimal hyper-parameter, and the final SVM model is obtained with the entire training set
Table~\ref{tab:SVM} shows the precision, recall, F1 score, support and Macro-average F-Score of the trained classifier for each topic category.
Then, the obtained SVM model classifies the rest of topics. Table~\ref{SVM_res} shows the number of topics of each category for each country and region.

The categories with higher percentages are topics of Wuhan \& China and policy and daily life. In general, the number of topics about policy and daily life is much higher in Luxembourg ($56.6\%$) than in other countries ($ave = 33.0\%$). France, on the other hand, shows a high level of interest in local news ($30.2\%$), compared with other countries ($9.4\%$). In terms of the overall data of GR, however, it does not show particular differences compared with other countries. Note that as there may be cases where the cluster for a topic contains no more than two tweets, we treat such topics as the invalid topic and remove them. This leads to a different total number of topics in each country. Next, we introduce dates to plot the changes in categories over time.

\begin{table}[!h]
\centering
\begin{tabular}{|r|r|r|r|r|}
\hline
\multicolumn{1}{|l|}{Category} & \multicolumn{1}{l|}{Precision} & \multicolumn{1}{l|}{Recall} & \multicolumn{1}{l|}{F1-score} & \multicolumn{1}{l|}{support}\\ \hline
1                               & 0.89                           & 0.77                        & 0.82 &163                          \\ \hline
2                               & 0.92                           & 0.93                        & 0.93 & 166                          \\ \hline
3                               & 0.80                           & 0.79                        & 0.80 &155                          \\ \hline
4                               & 0.74                           & 0.86                        & 0.80 &155                          \\ \hline
5                               & 0.73                           & 0.68                        & 0.71&149                          \\ \hline
6                               & 0.99                           & 1.00                        & 0.99&157                          \\ \hline
7                               & 0.97                           & 1.00                        & 0.98&142                          \\ \hline
\multicolumn{1}{|l|}{Macro avg} & 0.86                           & 0.86                        & 0.86&1,087                          \\ \hline
\end{tabular}
\centering
\caption{Metrics of the classification results}

\label{tab:SVM}
\end{table}

\begin{table}[!h]
\centering
\begin{tabular}{|r|r|r|r|r|r|r|}
\hline
\multicolumn{1}{|l|}{Category} & \multicolumn{1}{l|}{~~~GR~~} & \multicolumn{1}{l|}{Luxembourg} & \multicolumn{1}{l|}{Belgium} & \multicolumn{1}{l|}{France} & \multicolumn{1}{l|}{Germany} & \multicolumn{1}{l|}{Total} \\ \hline
1 & 245 & 168 & 287 & 202 & 315 & 1,217 \\ \hline
2 & 64 & 34 & 48 & 65 & 41 & 252 \\ \hline
3 & 99 & 44 & 109 & 285 & 110 & 647 \\ \hline
4 & 134 & 77 & 114 & 52 & 167 & 544 \\ \hline
5 & 353 & 525 & 370 & 250 & 295 & 1,793 \\ \hline
6 & 23 & 7 & 23 & 31 & 15 & 99 \\ \hline
7 & 41 & 72 & 15 & 60 & 23 & 211 \\ \hline
\multicolumn{1}{|l|}{Total} & 959 & 927 & 966 & 945 & 966 & 4,763 \\ \hline
\end{tabular}%
\caption{Topic volume for each category/country (region)}
\label{SVM_res}
\end{table}

\begin{figure}[!t]
\centering
\includegraphics[width=1.0\textwidth]{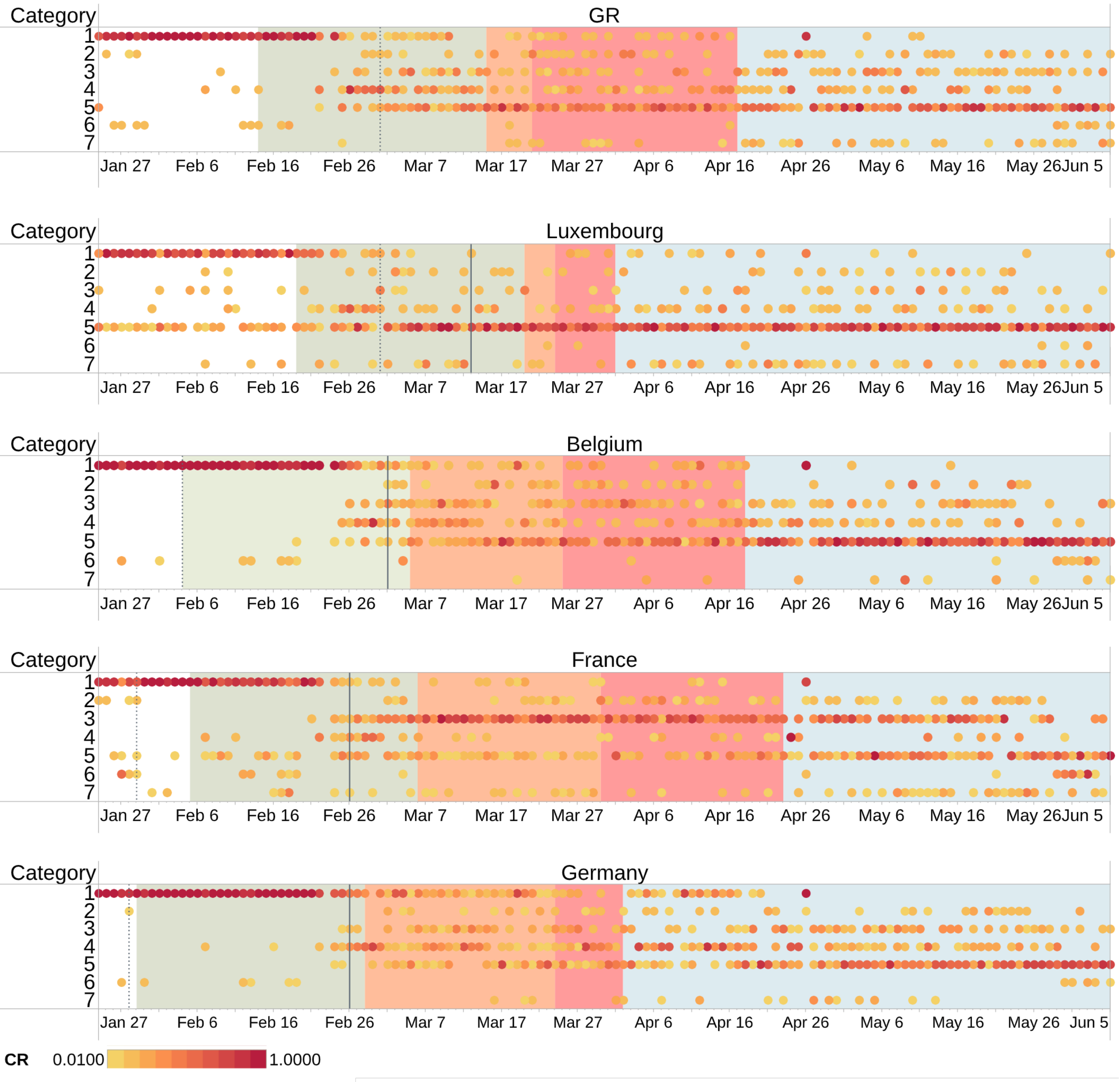}
\caption{Topic categories in GR and related countries}
\label{fig:Topic_clf}
\begin{tablenotes}
        \footnotesize
        \item[*] 1: Wuhan\& China 2: Measures 3: Local news 4: International news 5: Policy and daily life 6: Racism 7: Other
\end{tablenotes}
      
\end{figure}

Figure~\ref{fig:Topic_clf} shows the tweet volume contained in each category demonstrated in the form of percentage of the total tweet volume on that day (CR), with the darker red representing higher CR. The interval colored in white represents the period from 22 January to the start of {\it Pre-peak} period, other regions in different colours indicate, in sequence, {\it Pre-peak} period, {\it Free-contagious} period, {\it Measures} period, and {\it Decay} period. The black dotted line illustrates the date on which the first case appeared. The figure shows an interval between the date of the first case and the date of consecutive cases every day in GR. The solid black line indicates the date that new cases appear every day since that date. For ease of discussion, we name the day as `outbreak day' (OD).

+
\subsection{Research question {\sf RQ2}}
In this section, we aim to answer {\sf RQ2}, i.e., whether there are distinctive characteristics of these region and countries' topics about COVID-19 on Twitter, and whether GR, as a relational city, embodies any characteristics in the topics.

Figure~\ref{fig:Topic_clf} reveals that in France, Germany and Belgium, the appearance of the first case trigger only a small amount of discussions about the protective measures, and related discussions do not start to increase until OD. In other words, the public concerns in these region and countries do not really heed the protective measures until OD, when the virus is already spreading. This finding is at odds with the conclusion of Bento et al.~\cite{bento2020evidence} that the announcements of the first case have the greatest impact on the public concerns for searching basic information about COVID-19 and its symptoms.

Moreover, the report of first case does not stimulate discussions about policies and daily life as well, and discussions about it do not emerge frequently until OD.
This may be explained by the existence of a large interval between the date of the first case and OD (27.3 days on average) in France, Germany, and Belgium. During this interval, sporadic cases may not attract enough public concerns, and the public’s concerns is still focused on China-related news.

\begin{figure}[!t]
\subfigure[Policy and daily life]
{
    \begin{minipage}[b]{.49\linewidth}
        \centering
        \includegraphics[scale=0.4]{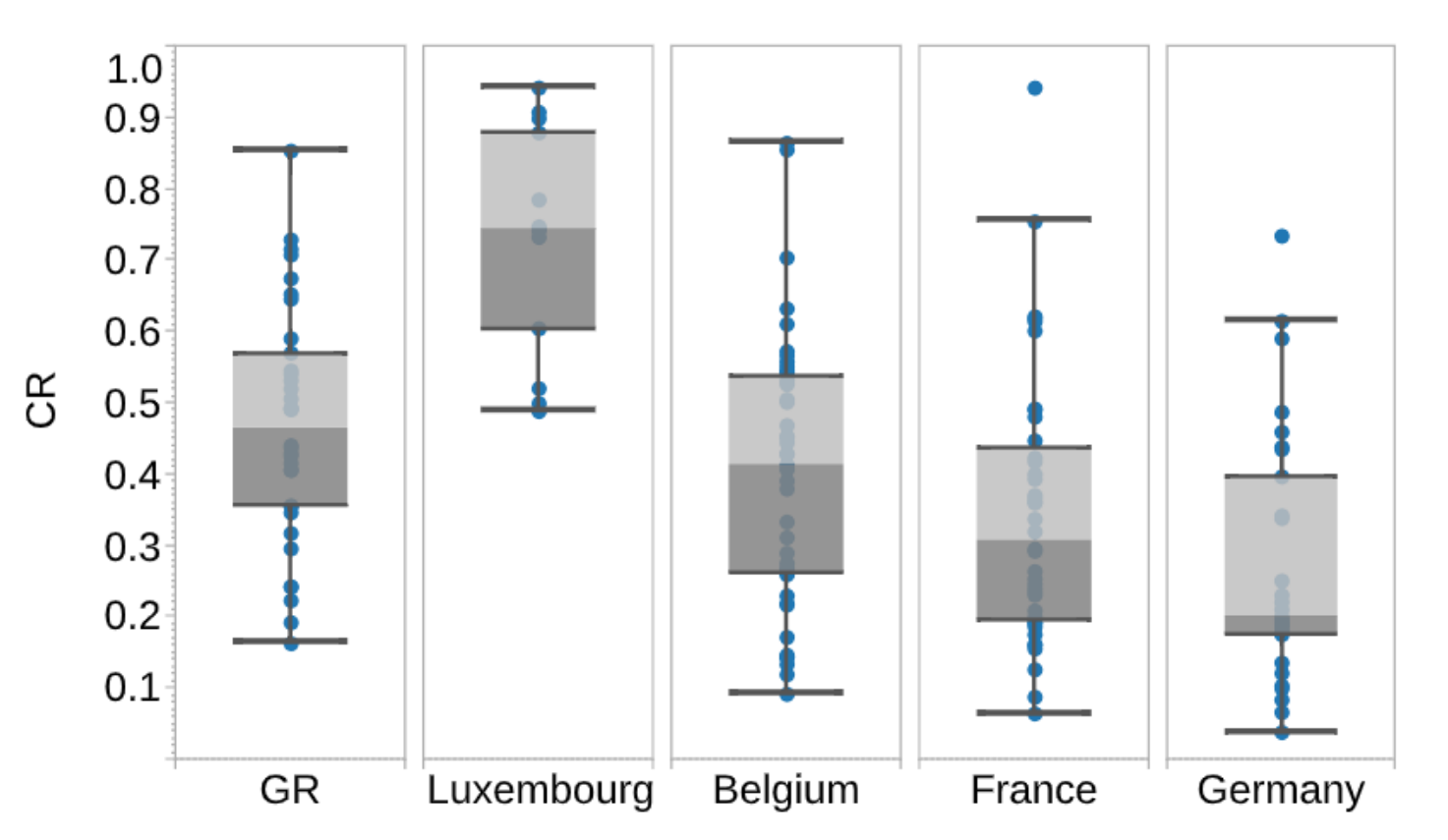}
        \label{fig_policy}
        \end{minipage}
}
\subfigure[Local news]
{
 	\begin{minipage}[b]{.49\linewidth}
        \centering
        \includegraphics[scale=0.4]{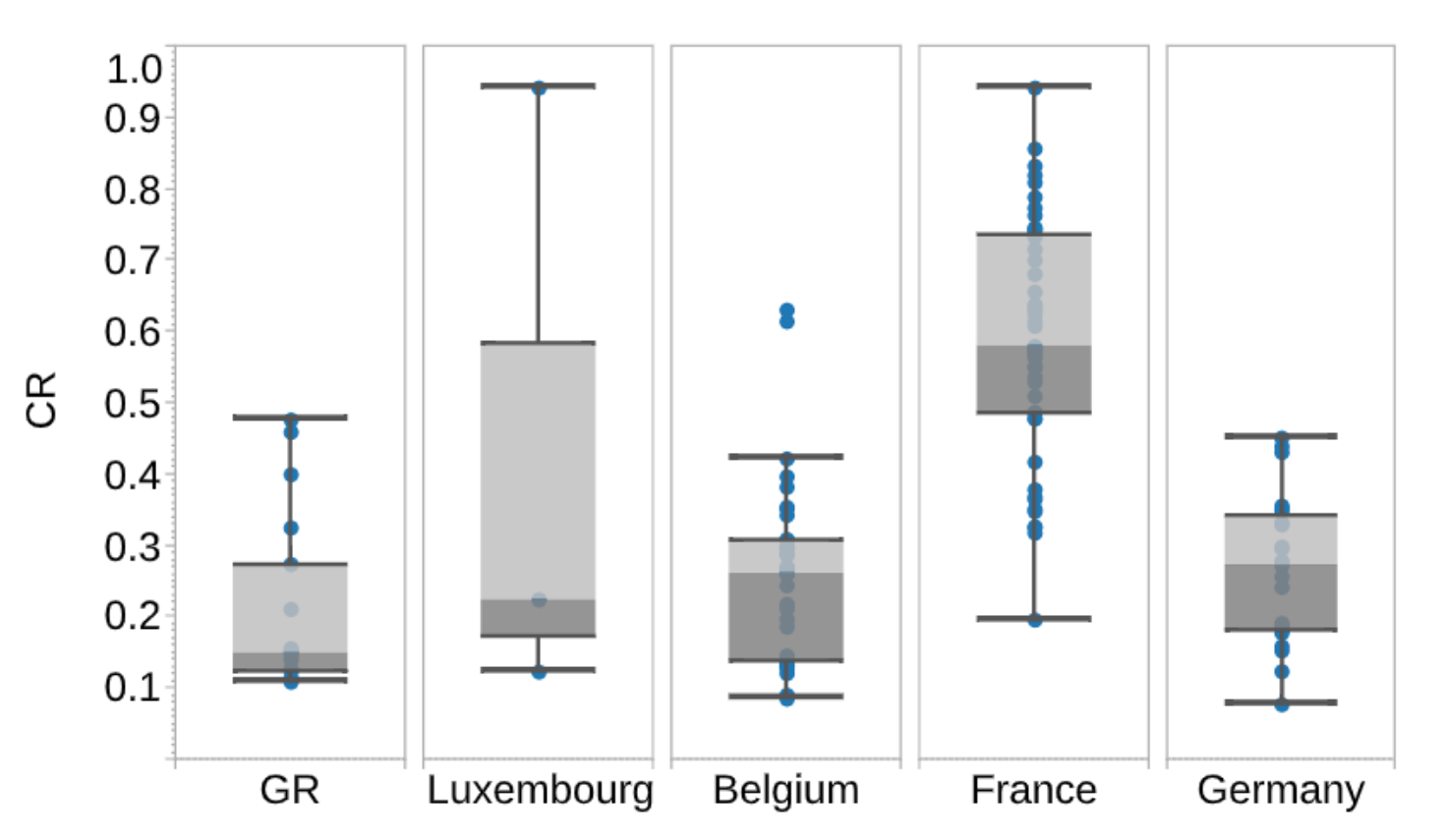}
        \label{fig_local}
    \end{minipage}
}

\caption{Distribution of proportion of tweets on `policy and daily life' and `local news' during Free-contagious and Measures period}
\label{fig:boxpolt_5}
\end{figure}

The situation is different in GR, a relational city, and in Luxembourg, its centre. Figure~\ref{fig:Topic_clf} shows that the public in Luxembourg and GR start to have discussions about measures $1-2$ days before the first case appears. 
Furthermore, during the {\it Pre-peak} period, the CR of measures is much higher in GR ($3.41\%$) and Luxembourg ($7.62\%$) than in France ($1.90\%$), Belgium ($1.84\%$) and Germany ($0.0\%$). It should be noted that discussions of measures are not totally non-existent in Germany, but the tweet volume may be too small to be recognised as separate topics during the topic modelling process. By comparing the topics discussed in other countries of the same time, this may be explained by the late occurrence of the first case in Luxembourg and GR, where the other three countries have already passed OD, the outbreak in other countries may have attracted public concerns in GR and Luxembourg.
Concurrently, the results indicate that GR exhibits a high level of interest in policy and daily life with $47.1\%$ of total tweet volume during the {\it Free-contagious} and the {\it Measures} period, while for Luxembourg, this rate is $66.1\%$. Figure~\ref{fig_policy} shows boxplots of the distribution of the CR on policy and daily life during the {\it Free-contagious} and the {\it Measures} period. This shows that the public is more responsive to policies as a region that relies on foreign labour and has high mobility than Belgium, France and Germany.

The reason why {\it Free-contagious} is a period more transient in Luxembourg and GR compared with other regions is still unclear, but part of the reason may stem from the fact that the public concerns to the virus itself during {\it Pre-peak} period led to better responsiveness to the anti-contagion policies in these region and countries.

\begin{figure}[!t]
\centering
\includegraphics[width =0.6\textwidth]{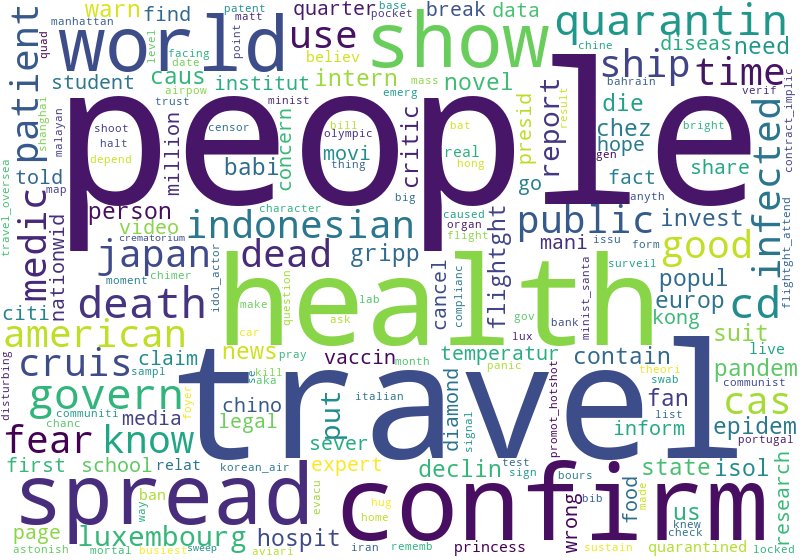}
\caption{Word cloud of Luxembourg Tweets from 2020-01-22 to 2020-03-01}
\label{fig:Lu_word}
\end{figure}

Interestingly, in Luxembourg, the discussion about policies and daily life persisted before the first case is announced and increased immediately after then. A word cloud of the topics from 22 January to 1 March (date of the first case) of Luxembourg is depicted in Figure~\ref{fig:Lu_word}, this shows that the topics are mainly travel-related. 
This may be explained by the fact that the proportion of foreign residents in the Luxembourg region is $47.4$\%\footnote{\url{https://bit.ly/3fdhgwj}}, and residents are more concerned about travel-related policies in Luxembourg and other countries. 

In addition, Figure~\ref{fig_local} illustrates that the {\it Free-contagious} and {\it Measures} periods coincided with the France municipal election, and thus the public concerns in local news among French is higher. In the end, during the {\it Decay} period, while there is a downward trend ($p <0.05$) in the total daily tweet volume, there is a upward trend ($p <0.05$) in the CR of policy and daily life, except in Luxembourg, where the rate is consistently high.

\section{Conclusion and Discussion}
\label{sec:conclusion}
In this paper, we studied the information related to COVID-19 on Twitter, introduced the concept of relational city and chose the Greater Region and its related country for our exploratory study. Our analysis has focused on two dimensions of pandemic information on Twitter, i.e., tweet volume and tweet text.

Based on the Spatio-temporal analysis of the correlation between tweet volume and COVID-19 cases during the four periods of the pandemic, our answer to {\sf RQ1} is that tweet volume and COVID-19 cases in GR and related countries are correlated, but this strong correlation only exists during the {\it Pre-peak} period of the pandemic. Regardless of the time at which $R(t)$ peaks, there is a 5-6 day lead between tweet volume and COVID-19 cases.

For the occurrence of this lead, we are tempted to consider whether this lead hinges on the incubation period of COVID-19.
Although the current research on the incubation period of COVID-19 is inconclusive, several studies have suggested that the incubation period of COVID-19 is on average 5-6 days~\cite{who_lag,lauer2020incubation,lei2020clinical}. In this regard, we speculate that the 5-6 day lead may be related to the lag between infection and the onset of symptoms to be detectable and confirmed.

For {\sf RQ2}, we found significant differences in the topics about COVID-19 concerned by Twitter users from different regions. While initially, the main topic is about China, the analysed region and countries reacted differently on the topic after the first case emerged. In France, Germany, and Belgium, the first case did not attract much attention to anti-contagious and treatment measures, policies and local news, until a complete outbreak. Whileas GR, as a relational city with a large number of cross-border workers, has shown high interest in policies since the first case, even if no lockdown policy has been implemented at that time. At the same time, Luxembourg, which has a foreign resident population of $47.4\%$, has shown a great concern for policies including travel from the beginning of the pandemic, which is not found in the other analysed countries.

We speculate that the reason for this can be explained by the fact that in these countries there was not an immediate outbreak of the pandemic after the first case, but rather after an interval of an average of 27.3 days. Thus, there may be an underestimation of the severity of the pandemic by the public in these countries during the {\it Pre-peak} period.
We tentatively suggest that a possible explanation for this phenomenon is \emph{optimism bias}, which makes people believe their exposure risk to disease is low~\cite{van2020using}. During a pandemic, people often exhibit an optimism bias, a cognitive bias that causes someone to believe that they will be less likely to get involved in negative events~\cite{sharot2011optimism}. Here, we offer a speculative interpretation that even though these countries have shown sustained and long-term concern about COVID-19 occurring in China on Twitter, optimism bias emerged when COVID-19 appeared, causing the public to ignore the emergence of the cases and to pay little attention on anti-contagious measures and government policies~\cite{paek2008public}. Further analysis of this issue will be undertaken in future studies.

Our results in the current paper can be used to understand topics being discussed on Twitter, and the differences exhibited in GR, the relational city, when facing the pandemic. At the same time, we make a speculative conclusion of the ideal point of time to conduct the pandemic precaution advocacy which help to provide policy support.

There are still some limitations of our study. 
First, in our dataset, we did not detect misleading information posted by bots, which can lead to a possible bias in topical modelling and classification.

For our initial exploration of topic categories, we chose SVM to build a baseline method for topic classification. We will utilise other state-of-the-art text classification methods to refine the classification in further study.
Second, our case study has some statistical limitations. Data from more countries will be included in future studies to ensure the statistical significance of the conclusions.

Third, more research can be performed based on our dataset. For example, in future, we will conduct sentiment analysis on the tweets of different categories at each pandemic period to find out the changing in the public’s sentiment about the pandemic and how it differs between GR and other countries. And for {\sf RQ2}, multi-class sentiment analysis with BERT will be conducted to figure out whether and to what extent people are optimistic or pessimistic about being affected by a pandemic during the {\it Pre-peak} period.
Finally, during the writing of this article, the second wave of COVID-19 emerges in Luxembourg and other studied countries.
In a future study, we will conduct a comparative study focusing on the regions that have the second wave. Sentiment analysis and text classification with the state-of-the-art method will be deployed to investigate whether OSNs information impact public attitude and behaviour.
We will attempt to identify topics that may help to predict the second wave, such as laxity or resistance to policies and anti-infection measures. Such timely indicators are potentially useful for making appropriate policy adjustments to avoid a new pandemic outbreak.

\medskip
\noindent{\bf Acknowledgements.}
This work was partially supported by Luxembourg’s Fonds National de la Recherche, via grant {\sf COVID-19/2020-1/14700602} (PandemicGR).

\bibliographystyle{elsarticle-num}
\bibliography{pandemic}

\begin{thebibliography}{10}
\expandafter\ifx\csname url\endcsname\relax
  \def\url#1{\texttt{#1}}\fi
\expandafter\ifx\csname urlprefix\endcsname\relax\def\urlprefix{URL }\fi
\expandafter\ifx\csname href\endcsname\relax
  \def\href#1#2{#2} \def\path#1{#1}\fi

\bibitem{cinelli2020covid}
M.~Cinelli, W.~Quattrociocchi, A.~Galeazzi, C.~M. Valensise, E.~Brugnoli, A.~L.
  Schmidt, P.~Zola, F.~Zollo, A.~Scala, {The covid-19 social media infodemic},
  arXiv preprint arXiv:2003.05004 (2020).

\bibitem{singh2020first}
L.~Singh, S.~Bansal, L.~Bode, C.~Budak, G.~Chi, K.~Kawintiranon, C.~Padden,
  R.~Vanarsdall, E.~Vraga, Y.~Wang, {A first look at COVID-19 information and
  misinformation sharing on Twitter}, arXiv preprint arXiv:2003.13907 (2020).

\bibitem{jahanbin2020using}
K.~Jahanbin, V.~Rahmanian, {Using Twitter and web news mining to predict
  COVID-19 outbreak}, Asian Pacific Journal of Tropical Medicine 13 (2020)
  26--28.

\bibitem{wang2011collaborative}
C.~Wang, B.~David~M., {Collaborative topic modelling for recommending
  scientific articles}, in: Proceedings of the 2011 International Conference on
  Knowledge Discovery and Data Mining (SIGKDD), ACM, 2011, pp. 448--456.

\bibitem{ordun2020exploratory}
C.~Ordun, S.~Purushotham, E.~Raff, {Exploratory analysis of covid-19 tweets
  using topic modelling, UMAP, and digraphs}, arXiv preprint arXiv:2005.03082
  (2020).

\bibitem{medford2020infodemic}
R.~J. Medford, S.~N. Saleh, A.~Sumarsono, T.~M. Perl, C.~U. Lehmann, {An
  "Infodemic": Leveraging high-volume Twitter data to understand public
  sentiment for the COVID-19 outbreak}, medRxiv (2020).

\bibitem{sharma2020covid}
K.~Sharma, S.~Seo, C.~Meng, S.~Rambhatla, Y.~Liu, {COVID-19 on Social Media:
  Analyzing Misinformation in Twitter Conversations}, arXiv preprint
  arXiv:2003.12309 (2020).

\bibitem{gupta2020tracking}
S.~Gupta, T.~D. Nguyen, F.~L. Rojas, S.~Raman, B.~Lee, A.~Bento, K.~I. Simon,
  C.~Wing, {Tracking public and private response to the COVID-19 epidemic:
  Evidence from state and local government actions}, Tech. rep., National
  Bureau of Economic Research (2020).

\bibitem{bento2020evidence}
A.~I. Bento, T.~Nguyen, C.~Wing, F.~Lozano-Rojas, Y.~Y. Ahn, K.~Simon,
  {Evidence from Internet search data shows information-seeking responses to
  news of local COVID-19 cases}, Proceedings of the National Academy of
  Sciences of the United States of America (PNAS) 117~(21) (2020).

\bibitem{lopez2020understanding}
C.~E. Lopez, M.~Vasu, C.~Gallemore, {Understanding the perception of COVID-19
  policies by mining a multilanguage Twitter dataset}, arXiv preprint
  arXiv:2003.10359 (2020).

\bibitem{thelwall2020retweeting}
M.~Thelwall, S.~Thelwall, {Retweeting for COVID-19: Consensus building,
  information sharing, dissent, and lockdown life}, arXiv preprint
  arXiv:2004.02793 (2020).

\bibitem{balcan2009multiscale}
D.~Balcan, V.~Colizza, B.~Gon{\c{c}}alves, H.~Hu, J.~J. Ramasco, A.~Vespignani,
  Multiscale mobility networks and the spatial spreading of infectious
  diseases, Proceedings of the National Academy of Sciences 106 (2009)
  21484--21489.

\bibitem{sigler2013relational}
T.~J. Sigler, Relational cities: Doha, panama city, and dubai as 21st century
  entrep{\^o}ts, Urban Geography 34 (2013) 612--633.

\bibitem{hesse2020relational}
M.~Hesse, M.~Rafferty, {Relational cities disrupted: reflections on the
  particular geographies of COVID-19 For small but global urbanisation in
  Dublin, Ireland, and Luxembourg City, Luxembourg}, Tijdschrift voor
  economische en sociale geografie 111~(3) (2020) 451--464.

\bibitem{heesterbeek1996concept}
J.~A.~P. Heesterbeek, K.~Dietz, {The concept of Ro in epidemic theory},
  Statistica Neerlandica 50~(1) (1996) 89--110.

\bibitem{devlin2018bert}
J.~Devlin, M.~W. Chang, K.~Lee, K.~Toutanova, {BERT: Pre-training of deep
  bidirectional transformers for language understanding}, in: Proceedings of
  the 2019 Annual Meeting of the Association for Computational Linguistics:
  Human Language Technologies (HLT), Vol.~1, ACL, 2019, pp. 4171--4186.

\bibitem{blei2003lda}
D.~M. Blei, A.~Y. Ng, M.~T.~I. Jordan, {Latent dirichlet allocation}, Journal
  of Machine Learning Research 3 (2003) 993--1022.

\bibitem{chang2011libsvm}
C.-C.~C. Chang, C.-J.~J. Lin, {LIBSVM: A library for support vector machines},
  Transactions on Intelligent Systems and Technology 2~(3) (2011) 1--27.

\bibitem{COVID-19Dataset}
E.~Chen, K.~Lerman, E.~Ferrara, {COVID-19: The First Public Coronavirus Twitter
  Dataset}, arXiv preprint arXiv:2003.07372 (2020).

\bibitem{st2012can}
C.~{St Louis}, G.~Zorlu, {Can Twitter predict disease outbreaks?}, British
  Medical Journal 344~(7861) (2012).

\bibitem{hsiang2020effect}
S.~Hsiang, D.~Allen, S.~Annan-Phan, K.~Bell, I.~Bolliger, T.~Chong,
  H.~Druckenmiller, L.~Y. Huang, A.~Hultgren, E.~Krasovich, Others, {The effect
  of large-scale anti-contagion policies on the COVID-19 pandemic}, Nature
  (2020) 1--9.

\bibitem{courtemanche2020strong}
C.~Courtemanche, J.~Garuccio, A.~Le, J.~Pinkston, A.~Yelowitz, {Strong social
  distancing measures in the United States reduced The COVID-19 Growth Rate},
  Health Affairs (2020) 10--1377.

\bibitem{dergiades2020effectiveness}
T.~Dergiades, C.~Milas, T.~Panagiotidis, {Effectiveness of government policies
  in response to the COVID-19 outbreak}, SSRN (2020).

\bibitem{hu2020more}
D.~Hu, X.~Lou, Z.~Xu, N.~Meng, Q.~Xie, M.~Zhang, Y.~Zou, J.~Liu, G.~P. Sun,
  F.~Wang, {More effective strategies are required to strengthen public
  awareness of COVID-19: Evidence from Google trends}, Journal of Global Health
  10~(1) (2020).

\bibitem{effenberger2020association}
M.~Effenberger, A.~Kronbichler, J.~I. Shin, G.~Mayer, H.~Tilg, P.~Perco,
  {Association of the COVID-19 pandemic with Internet search volumes: A Google
  trends$^{TM}$ Analysis} (2020).

\bibitem{joulin2016fasttext}
A.~Joulin, E.~Grave, P.~Bojanowski, M.~Douze, H.~J{\'{e}}gou, T.~Mikolov,
  {Fasttext. zip: Compressing text classification models}, arXiv preprint
  arXiv:1612.03651 (2016).

\bibitem{yan2013biterm}
X.~Yan, J.~Guo, Y.~Lan, X.~Cheng, {A biterm topic model for short texts}, in:
  Proceedings of the 22nd International Conference on World Wide Web, 2013, pp.
  1445--1456.

\bibitem{bettencourt2008real}
L.~M. L. M.~A. Bettencourt, R.~M. Ribeiro, {Real time Bayesian estimation of
  the epidemic potential of emerging infectious diseases}, PLoS One 3~(5)
  (2008) e2185.

\bibitem{younis2020social}
J.~Younis, H.~Freitag, J.~S. Ruthberg, J.~P. Romanes, C.~Nielsen, N.~Mehta,
  Social media as an early proxy for social distancing indicated by the
  covid-19 reproduction number: observational study, JMIR Public Health and
  Surveillance 6 (2020) e21340.

\bibitem{smith2016towards}
M.~C. Smith, D.~A. Broniatowski, M.~J. Paul, M.~Dredze, {Towards real-time
  measurement of public epidemic awareness: Monitoring influenza awareness
  through twitter}, in: Spring Symposium on Observational Studies Through
  Social Media and Other Human-generated Content, Vol. 20052, 2016, p. e198.

\bibitem{wagstaff2001constrained}
K.~Wagstaff, C.~Cardie, S.~Rogers, S.~Schr{\"{o}}dl, Others, {Constrained
  k-means clustering with background knowledge}, in: Proceedings of the 2001
  International Conference on Machine Learning (ICML), Vol.~1, Citeseer, 2001,
  pp. 577--584.

\bibitem{mcinnes2018umap}
L.~McInnes, J.~Healy, J.~Melville, {Umap: Uniform manifold approximation and
  projection for dimension reduction}, arXiv preprint arXiv:1802.03426 (2018).

\bibitem{gutierrez2016detecting}
E.~D. Guti{\'e}rrez, E.~Shutova, P.~Lichtenstein, G.~de~Melo, L.~Gilardi,
  Detecting cross-cultural differences using a multilingual topic model,
  Transactions of the Association for Computational Linguistics 4 (2016)
  47--60.

\bibitem{ramos2003using}
J.~Ramos, Others, {Using TF-IDF to determine word relevance in document
  queries}, in: Proceedings of the 2003 instructional conference on machine
  learning (ICML), Vol. 242, 2003, pp. 133--142.

\bibitem{o2015analysis}
D.~O'callaghan, D.~Greene, J.~Carthy, P.~Cunningham, D.~O'Callaghan, D.~Greene,
  J.~Carthy, P.~Cunningham, {An analysis of the coherence of descriptors in
  topic modelling}, Expert Systems with Applications 42~(13) (2015) 5645--5657.

\bibitem{newman2010automatic}
D.~Newman, J.~H. Lau, K.~Grieser, T.~Baldwin, {Automatic evaluation of topic
  coherence}, in: Proceedings of the 2010 Annual Meeting of the Association for
  Computational Linguistics: Human Language Technologies, ACL, 2010, pp.
  100--108.

\bibitem{aranganayagi2007clustering}
S.~Aranganayagi, K.~Thangavel, {Clustering categorical data using silhouette
  coefficient as a relocating measure}, in: Proceedings of the 2007
  International Conference on Computational Intelligence and Multimedia
  Applications, Vol.~2, 2008, pp. 13--17.

\bibitem{abd2020top}
A.~Abd-Alrazaq, D.~Alhuwail, M.~Househ, M.~Hamdi, Z.~Shah, Top concerns of
  tweeters during the covid-19 pandemic: infoveillance study, Journal of
  medical Internet research 22~(4) (2020).

\bibitem{chawla2002smote}
N.~V. Chawla, K.~W. Bowyer, L.~O. Hall, W.~P. Kegelmeyer, {SMOTE: synthetic
  minority over-sampling technique}, Journal of Artificial Intelligence
  Research 16 (2002) 321--357.

\bibitem{who_lag}
{World Health Organization}, {Coronavirus disease 2019 (COVID-19) Situation
  Report – 73}, Tech. rep., World Health Organization (2020).

\bibitem{lauer2020incubation}
S.~A. Lauer, K.~H. Grantz, Q.~Bi, F.~K. Jones, Q.~Zheng, H.~R. Meredith, A.~S.
  Azman, N.~G. Reich, J.~Lessler, {The incubation period of Coronavirus disease
  2019 (COVID-19) from publicly reported confirmed cases: Estimation and
  application}, Annals of Internal Medicine 172~(9) (2020) 577--582.

\bibitem{lei2020clinical}
S.~Lei, F.~Jiang, W.~Su, C.~Chen, J.~Chen, W.~Mei, L.-Y. Zhan, Y.~Jia,
  L.~Zhang, D.~Liu, Others, {Clinical characteristics and outcomes of patients
  undergoing surgeries during the incubation period of COVID-19 infection},
  EClinicalMedicine (2020) 100331.

\bibitem{van2020using}
M.~C. Smith, D.~A. Broniatowski, M.~J. Paul, M.~Dredze, {Using social and
  behavioural science to support COVID-19 pandemic response}, Nature Human
  Behaviour 4~(5) (2020) 1--12.

\bibitem{sharot2011optimism}
T.~Sharot, {The optimism bias}, Current Biology 21~(23) (2011) 941--945.

\bibitem{paek2008public}
H.-J.~J. Paek, K.~Hilyard, V.~S. Freimuth, J.~K. Barge, M.~Mindlin, {Public
  support for government actions during a flu pandemic: lessons learned from a
  statewide survey}, Health Promotion Practice 9~(4) (2008) 60--72.

\end{thebibliography}

\end{document}